\documentclass[a4paper,11pt]{article}
\pdfoutput=1

\usepackage[T1]{fontenc} 
\usepackage[utf8]{inputenc}
\usepackage{physics}
\usepackage{xcolor}
\usepackage{mathrsfs}
\usepackage{tikz}
\usepackage{colortbl}
\usepackage{amsmath,amssymb,amsbsy,amstext}
\usepackage{svg}

\usepackage{hyperref}
\hypersetup{colorlinks=true,urlcolor=pacificblue,citecolor=blue,linkcolor=pacificblue}

\numberwithin{equation}{section}
\graphicspath{{figures/}}

\definecolor{pacificblue}{cmyk}{0.95,0.3,0, 0.5}

\usepackage[framemethod=TikZ]{mdframed}
\mdfsetup{roundcorner=5pt}

\newmdenv[
skipabove=5pt,
skipbelow=7pt,
rightline=false,
leftline=false,
topline=false,
bottomline=false,
backgroundcolor=gray!30,
innerleftmargin=5pt,
innerrightmargin=5pt,
innertopmargin=5pt,
innerbottommargin=5pt,
leftmargin=0cm,
rightmargin=0cm,
linewidth=4pt]{eBox}

\newcommand{\eq}[1]{\begin{equation}#1\end{equation}}
\newcommand{\eqa}[1]{\begin{align}#1\end{align}}
\newcommand{\spl}[1]{\begin{split} #1 \end{split}}
\newcommand{\fg}[1]{\begin{figure}[tbp]\centering #1 \end{figure}}
\newcommand{\p}{\partial}

\newcommand{\Mpl}{M_{\rm Pl}}
\newcommand{\f}{f_a}
\newcommand{\F}{\tilde{F}}
\newcommand{\kv}{\vb{k}}
\newcommand{\kpv}{\vb{k'}}
\newcommand{\xv}{\vb{x}}

\newcommand{\vp}{\varphi}
\newcommand{\xr}{{\rm x}}
\newcommand{\vrk}{{\varsigma}}

\begin{document}

\begin{titlepage}

\setcounter{page}{1} \baselineskip=15.5pt \thispagestyle{empty}
\begin{flushright} {\footnotesize YITP-20-90, IPMU20-0078}  \end{flushright}
\vspace{5mm}
\vspace{0.5cm}
\def\thefootnote{\fnsymbol{footnote}}
\bigskip

\begin{center}
{\fontsize{22}{15}\selectfont  \bf 
Analytic study of dark photon and \\ \medskip gravitational wave production from axion
}
\end{center}
\vspace{0.5cm}

\begin{center}
{ Borna Salehian$^{1}\footnote{salehian@ipm.ir}$, Mohammad Ali Gorji$^{2}\footnote{gorji@yukawa.kyoto-u.ac.jp}$, Shinji Mukohyama$^{2,3}\footnote{shinji.mukohyama@yukawa.kyoto-u.ac.jp}$, Hassan Firouzjahi$^{1}\footnote{firouz@ipm.ir}$ }
\\[.7cm]

{\small \textit{$^{1}$School of Astronomy, 
Institute for Research in Fundamental Sciences (IPM) \\ 
 P.~O.~Box 19395-5531, Tehran, Iran }} \\

{\small \textit{$^{2}$Center for Gravitational Physics,
Yukawa Institute for Theoretical Physics \\
Kyoto University, 606-8502, Kyoto, Japan}} \\
{\small \textit{$^{3}$Kavli Institute for the Physics and Mathematics of the Universe (WPI), 
The University of Tokyo Institutes for Advanced Study, 
The University of Tokyo, Kashiwa, Chiba 277-8583, Japan}} \\
\end{center}  
\vspace{.7cm}

\hrule \vspace{0.3cm}
{\noindent \textbf{Abstract} \\[0.2cm]
\noindent
Axion-like fields heavier than about $10^{-27}$eV are expected to oscillate in the radiation dominated epoch when the Hubble parameter drops below their mass. Considering the Chern-Simons coupling with a dark gauge boson, large amount of dark photons are produced during a short time interval through tachyonic resonance instability. The produced dark photons then source gravitational tensor modes leading to chiral gravitational waves. Through this process, one can indirectly probe a large parameter space of coupled axion-dark photon models. In this work we first find an analytic expression for the number density of the dark photons produced during the tachyonic resonance regime. Second, by using the saddle point approximation we find an analytic expression for the gravitational wave spectrum in terms of the mass, coupling and misalignment angle. Our analytic results can be used for the observational analysis of these types of scenarios.
\vspace{0.3cm} \hrule}

\end{titlepage}

\setcounter{footnote}{0}

\section{Introduction}\label{eq:intro}

Most of the energy content of the universe is in the dark sector and, nowadays, the so-called dark energy and dark matter problems are the big challenges for the standard model of cosmology. It is then reasonable to assume that the dark sector, analogous to the ordinary visible matter sector, has a rich structure of its own particles and forces such as (pseudo-)scalars and gauge bosons \cite{Essig:2013lka}. 
 
Axion is one of the well-motivated examples of pseudo-scalar fields in the dark sector. While from the bottom-up point of view it can be a solution to the strong CP problem in the standard model of particle physics, from the top-down viewpoint it can arise in UV complete theories, such as string theory, in which a global Peccei-Quinn symmetry is spontaneously broken \cite{Kim:1986ax,Ringwald:2014vqa,Witten:1984dg,Arvanitaki:2009fg}. The properties of an axion are characterized by two independent parameters: the scale $f_a$ of symmetry breaking and the mass $m$ which arises due to nonperturbative effects. There are different bounds on the coupling and mass of axions. For example, avoiding the cosmological overabundance of the QCD axion puts an upper bound $f_a\lesssim10^{12}$GeV. However, bounds on the coupling can be relaxed by considering a coupling of the axion to hidden photons or monopoles \cite{Agrawal:2017eqm,Kitajima:2017peg,Kawasaki:2015lpf} as such a coupling is predicted in some models based on string theory \cite{Conlon:2006tq}, late time entropy production before big bang nucleosynthesis \cite{Kawasaki:1995vt,Banks:1996ea,Dienes:2012jb}, and dynamical axion misalignment \cite{Dvali:1995ce,Graham:2018jyp,Guth:2018hsa,Co:2018phi,Buen-Abad:2019uoc}. Having larger values of $f_a$ makes it possible to consider scenarios in which axion-like fields play significant roles in the earlier times in the expansion history of the universe  which was the subject of many studies \cite{Anber:2006xt,Sorbo:2011rz,Barnaby:2012xt,Anber:2012du,Adshead:2016iae,Domcke:2016bkh,Choi:2018dqr}. The important assumption we make in the following is that the symmetry breaking happens before the end of inflation so that the axion acquires a homogeneous background produced by the misalignment mechanism \cite{Sikivie:2006ni}. 

While for the case of the QCD axion the mass is determined by the symmetry breaking scale (see Ref.~\cite{diCortona:2015ldu}) it is generally expected to be an independent parameter within a wide range of scales. When the Hubble expansion rate drops and becomes comparable to the axion mass during the expansion of the universe, the homogeneous background of axion starts to oscillate. In our scenario, for concreteness, we assume that $m\gtrsim10^{-27}$eV which corresponds to the Hubble scale at matter radiation equality \cite{Aghanim:2018eyx} such that the axion oscillations take place in  the radiation dominated (RD) universe. However, our analysis will be more or less the same for axions with smaller masses that start oscillation later, i.e. in the matter dominated era. Furthermore, we ignore the temperature dependence of the mass as well as the self-interactions in the axion field which does not make any qualitative differences in what follows. 

Oscillation of the axion results in an energy transfer to fields that are coupled to it via the mechanism of  parametric  resonance very similar to  the mechanism of preheating after inflation \cite{Kofman:1994rk,Kofman:1997yn}. The amplification of the fields coupled to the oscillating axion-like fields leads to some interesting features which were the subject of many recent investigations \cite{Hertzberg:2020dbk,Sun:2020gem,Wang:2020zur,Chu:2020iil,Kitajima:2018zco,Fukunaga:2019unq,Hertzberg:2018zte,Soda:2017sce,Yoshida:2017cjl}. In this work, we consider a $U(1)$ gauge boson in the dark sector coupled to the axion field. A dark gauge boson may arise in string compactifications \cite{Pospelov:2008zw,Goodsell:2010ie,Ringwald:2012hr} as well as in vector dark matter models \cite{Agrawal:2018vin,Dror:2018pdh}. The natural interaction between the axion and dark photon is the so-called Chern-Simons coupling. Such a coupling is very well studied in different contexts like magnetogenesis \cite{Fujita:2015iga,Adshead:2016iae,Okano:2020uyr,Garretson:1992vt}, inflation \cite{Anber:2009ua,Sorbo:2011rz,Mukohyama:2014gba}, preheating \cite{Adshead:2015pva} and in axion dark matter models to solve the problem of the overabundance of the axions \cite{Agrawal:2017eqm,Kitajima:2017peg}. The Lagrangian density for the dark sector takes the form
\begin{equation}\label{L-axion}
{\cal L} = - \frac{1}{2}(\p\phi)^2 - \frac{1}{2}m^2\phi^2
- \frac{1}{4}F_{\mu\nu} F^{\mu\nu} - \frac{{\alpha}_{a}}{4\f}\phi F_{\mu\nu} \F^{\mu\nu} \,,
\end{equation}   
where $\phi$ is the axion field, $F_{\mu\nu}$ is the field strength tensor of the dark photon with the $U(1)$ gauge symmetry, and ${\tilde F}^{\mu\nu} = \frac{1}{2} \epsilon^{\mu\nu\alpha\beta} F_{\alpha\beta}$ is the dual field strength tensor with $\epsilon^{\mu\nu\alpha\beta}$ being the Levi-Civita tensor associated to the metric with the convention $\epsilon^{0123}=1/\sqrt{g}$. The coupling to the dark photon is characterized by $\f$ and a dimensionless and model-dependent numerical factor ${\alpha}_{a}$. The latter can be designed to take large values (even $\order{100}$) for example in clockwork models \cite{Higaki:2015jag,Kaplan:2015fuy,Agrawal:2017cmd}. In this work, we are especially interested in large values ${\alpha}_{a}$ where, as we will see in more detail below, the dark photons become tachyonic and the particle production is very efficient. However, since the amplitude of axion oscillations decreases due to both expansion of the universe and energy transfer to photons, particle creation will soon stop being efficient. The whole process starts and ends very fast so that it takes place entirely in the RD universe.

While the axion field interacts weakly with standard model particles, most of the attempts towards detection of the axion is based on its coupling to the standard model (see Refs.~\cite{Graham:2015ouw,TheMADMAXWorkingGroup:2016hpc,Du:2018uak,Lawson:2019brd,Zarei:2019sva,Aprile:2020tmw}). However, the natural probe of the structure of the dark sector is detection of the gravitational wave (GW) signal. The reason is that according to the equivalence principle, any types of energy or matter would at least minimally couple to gravity. As a result, in the coupled axion-dark photon model described above, it is important to investigate the GW signal from huge amount of dark photons produced via tachyonic instability. This is the main idea investigated in the pioneering works of Refs.~\cite{Machado:2018nqk,Machado:2019xuc} in which the authors explored the dark photon production and the corresponding distinct signal in the GW spectrum mostly based on numerical analysis. According to their results, there is a good chance to detect the GW signal by the future experiments for a wide range of parameters \cite{Machado:2019xuc}. 

Note that the coupled system of axion and dark photon in an expanding background is very complicated. Fig.~\ref{fig:diagram} shows the interplay between different degrees of freedom in the system under consideration. Here is a brief qualitative description of the system. Let us for the moment ignore the metric fluctuations. The initial energy stored in the homogeneous part of the axion field is transferred to its inhomogeneous part as well as the dark photons due to the parametric resonance. The former is due to the self-coupling\footnote{In this work we have neglected the self-coupling and as written in Eq.~\eqref{L-axion} we have only considered the mass term.} of the axion potential while the latter is because of the Chern-Simons coupling. As the universe expands the energy transfer becomes less efficient. In the reverse order, the presence of dark photons and axion fluctuations affect the dynamics of the background axion. The former is usually called the backreaction effect. There is also energy transfer between dark photon and axion particles which is sometimes called backscattering. Furthermore, one must consider the fluctuations in the metric and their backreaction on the other fields' dynamics. The complicated and nonlinear dynamics described above is the main reason why most of the analyses performed in the literature are based on the numerical methods.
    
The aim of this paper is to study the model mostly with analytical techniques. Indeed, one should not expect to be able to obtain analytic understanding of the model in the nonlinear regime due to the complicated dynamics. Instead, we focus on the two legs in Fig.~\ref{fig:diagram} specified by bold orange arrows, that are the production of dark photons from background axion oscillations and also the tensor modes from the dark photons. For the case of dark photon production we use the method of successive scattering matrices used in preheating scenarios \cite{Kofman:1997yn} to find an analytic expression for the number of dark photons. We will try to find an estimate of the time when backreaction and backscattering effects become important and the analytic results cannot be trusted anymore. As for the GW production, we use our understanding from dark photon production to find an analytic expression for its spectrum. We obtain an approximate closed form expression for the spectrum of GWs by using the method of saddle point approximation as well as by using the fact that dark photons are mostly produced at a specific momentum. These analytic results, though being approximate, not only are useful in giving insights but also improve the template of the GW spectrum for the observational analysis \cite{Machado:2019xuc}.

\fg{
	\centering
	\includegraphics[width=0.4\textwidth]{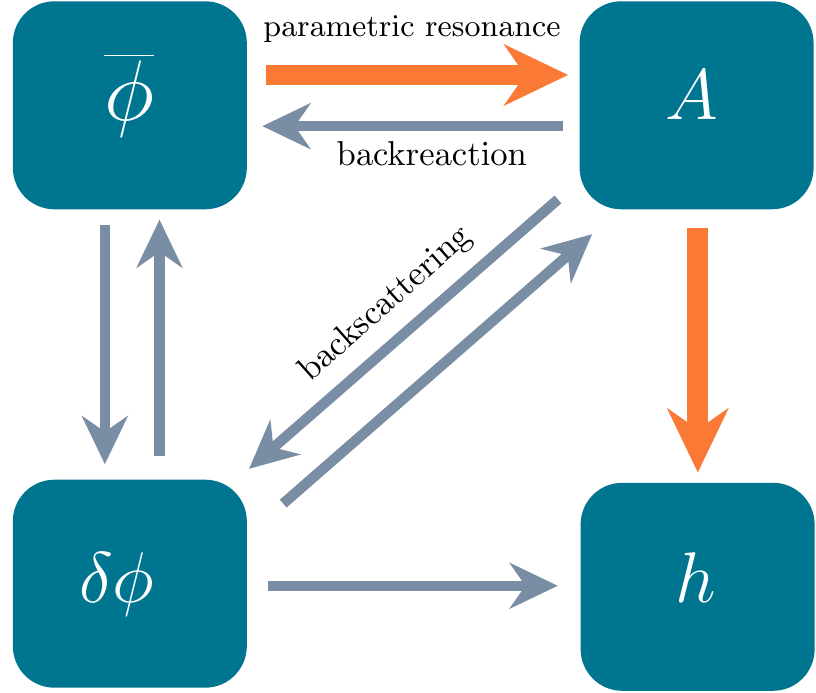}
	\caption{\small Schematic diagram showing the interplay between different degrees of freedom in the model. Ignoring metric perturbations we have axion background ($\bar{\phi}$), dark photons ($A$) and axion fluctuations ($\delta\phi$). Because of the axion oscillations, its own fluctuations and dark photons are produced due to parametric resonance. However, the dynamics is nonlinear as $A$ and $\delta\phi$ affect each other's dynamics (backscattering) as well as $\bar{\phi}$ (backreaction). Besides, we must include metric perturbations. Here we have shown only the tensor modes ($h$) which can be produced from photons and axion particles. The analytic investigations of this work is focused on the orange arrows.}
	\label{fig:diagram}
}

The rest of the paper is organized as follows. In section \ref{sec-vector-field}, based on the method previously used in the investigation of preheating scenarios after inflation, we find analytic solutions for the mode functions of the dark photons during the tachyonic and semi-tachyonic regimes and then we find an analytical expression for the number density of the produced dark photons. We also find an analytic expression for the correction to the effective number of relativistic degrees of freedom induced by the produced dark photons. In section \ref{sec-GWs}, we analytically compute the GW spectrum. Section \ref{sec-conclusion} is devoted to a summary of the results and conclusions. In appendices \ref{app:pol}, \ref{app:floquet}, and \ref{semitach}, we present some useful formulas which are used in the paper while in appendix \ref{app:perturbative-decay} we present an analytic study for the possibility of the perturbative decay of the axion. 

\section{Production of dark photons}\label{sec-vector-field}
In this section we explore non-perturbative production of the dark photons caused by the oscillations of the axion field $\phi$. As we explained in the Introduction, the model is defined by the Lagrangian density \eqref{L-axion} for the dark sector that is minimally coupled to the standard Einstein-Hilbert action. Varying the corresponding total action with respect to the axion field $\phi$ we find
\begin{equation}\label{eq:EOM-axion}
-\frac{1}{\sqrt{g}}\p_\mu(\sqrt{g}\p^\mu\phi)+m^2 \phi 
= -\frac{{\alpha}_{a}}{4\f}F_{\mu\nu}\F^{\mu\nu} \,,
\end{equation}
where the field strength associated to the gauge field $A_{\mu}$ is $F_{\mu\nu} = \partial_{\mu}A_{\nu} - \partial_{\nu}A_{\mu}$. Taking the variation of the action with respect to the gauge field, we find the corresponding equation of motion
\begin{equation}\label{eq:motEM}
\frac{1}{\sqrt{g}}\p_\mu(\sqrt{g}F^{\mu\nu})=-\frac{{\alpha}_{a}}{\f}(\p_\mu\phi)\F^{\mu\nu}\,,
\end{equation}
where we have used the identity $\frac{1}{\sqrt{g}}\p_\mu(\sqrt{g}\F^{\mu\nu})=0$. With the assumed range of the mass of the axion $m\gtrsim10^{-27}$eV, oscillation starts in the RD era when the Hubble expansion rate drops below the axion mass scale. Moreover, the process of non-perturbative dark photon production and the emission of  GW are very short and fits entirely into the RD epoch. The energy density of the universe is then dominated by its radiation content and any contribution from the dark sector can be ignored. As a result, from the Friedmann equation we find that the background geometry is 
\begin{equation}\label{metric}
ds^2 = - dt^2 + a(t)^2 \delta_{ij}dx^i dx^j = a(\eta)^2 \big(-\dd{\eta}^2+ \delta_{ij}dx^i dx^j \big) \,,
\end{equation}
where $t$ is the cosmic time, $\eta=\int dt/a(t)$ is the conformal time and $a \sim t^{1/2} \sim\eta$ is the scale factor. 

The electric and magnetic fields associated to the gauge field are defined from the field strength tensor as $F_{0i}= -a E_i$ and $F_{ij} = a^{2}\epsilon_{ijk}B_k$	 such that they coincide with those measured by an inertial observer\footnote{This definition can be deduced by looking at the components of the strength tensor in the tetrad basis where the metric takes the form $g_{\mu\nu} = \eta_{ab} e^a{}_\mu e^b{}_\nu$ with $\eta_{ab}$ being the local inertial metric and the spacetime curvature is encoded in tetrads $e^a{}_\mu$. From Eq.~\eqref{metric}, the nonzero components of the tetrads are $e^t{}_t = 1$ and $e^{a}{}_i=a(t)\delta^a_i$. Expanding the strength tensor field in the tetrad basis $F_{\mu\nu}={\mathrm f}_{ab}e^a{}_\mu e^b{}_\nu$, we find that $F_{ti} = a {\mathrm f}_{ta}\delta^a_i$ and $F_{ij} = a^2 {\mathrm f}_{ab} \delta^a_i \delta^b_j$. We then define the electric and magnetic fields with respect to the flat space counterpart of the strength tensor ${\mathrm f}_{ab}$ as $E_i\equiv - {\mathrm f}_{ta}\delta^a_i$ and $B_i\equiv \frac{1}{2} \epsilon_{i}{}^{jk}{\mathrm f}_{ab}\delta^a_j\delta^b_k$.}. The equation of motion for the axion field $\phi$ \eqref{eq:EOM-axion} in the background (\ref{metric}) takes the form
\eq{
	\label{eq:phi}
	\ddot{\phi}+3H\dot{\phi}-\frac{1}{a^2}\p^2\phi+m^2\phi=\frac{{\alpha}_{a}}{\f}E_iB_i\,,
}
where a dot denotes derivative with respect to the cosmic time and $H=\dot{a}/a$ is the Hubble expansion rate. Further, we impose the Coulomb gauge condition $\delta^{ij}\p_iA_j=0$ to remove redundant degrees of freedom. Then, from Eq.~\eqref{eq:motEM} we get a constraint for $A_0$
\eq{
\p^2A_0=-a\frac{{\alpha}_{a}}{\f}B_i\p_i\phi\,,
}
with the solution
\eq{
	\label{eq:A0}
A_0=\frac{a{\alpha}_{a}}{4\pi\f}\int\dd[3]{x'}\frac{B_i\p_i'\phi(x')}{|x-x'|}\,.
}
Note that if we neglect spatial variation of the field $\phi$ then we get $A_0=0$ which means that the Coulomb gauge condition also implies the temporal gauge condition at leading order \cite{Adshead:2015pva}. The equation for the spatial part of the gauge field can be obtained from Eq. (\ref{eq:motEM}) which turns out to be
\eq{
\label{eq:Ai}
	\ddot{A}_i+H\dot{A}_i-\frac{1}{a^2}\p^2A_i
	+ \frac{{\alpha}_{a}}{a\f}\epsilon_{ijk} (\p_j \phi \dot{A}_k - \dot{\phi} \p_jA_k)
	= \frac{1}{a}\p_i\p_t(a{A}_0) + \frac{{\alpha}_{a}}{a\f}\epsilon_{ijk}\p_j\phi \p_kA_0 \,,
}
where $A_0$ is given by Eq.~\eqref{eq:A0}. As mentioned before, we assume that the Peccei-Quinn symmetry breaking happened before the end of inflation such that $\phi$ has a homogeneous expectation value denoted by $\bar{\phi}$ and in the following we ignore perturbations of the axion field. Note that if we consider self-interactions of the axion field or backscattering effects, perturbations of $\phi$ will be produced (see Ref.~\cite{Kofman:1994rk}). We will ignore these effects in our analysis. The equation for the homogeneous vacuum expectation value of the axion $\bar{\phi}$ then reads
\eq{
\label{eq:phibar}
		\ddot{\bar{\phi}}+3H\dot{\bar{\phi}}+m^2\bar{\phi}=\frac{{\alpha}_{a}}{\f}(E_iB_i)_0\,,
}     
where the subscript $0$ in the right hand side shows the zero mode of the electromagnetic source. Initially the energy density of the dark photons is assumed to be small and we can ignore the source. When the Hubble friction term becomes subdominant due to the expansion, the homogeneous background of the axion starts to oscillate. In this regime, the approximate solution is of the form
\begin{equation}\label{axion-osclliation}
\bar{\phi}(t)=\frac{\phi_{\rm os}}{a^{3/2}}\cos(mt)\,,
\end{equation}
where $\phi_{\rm os}$ is the initial amplitude at the time of the beginning of the oscillation $t_{\rm os}$ and for simplicity we have set $a(t_{\rm os})=1$ by rescaling of spatial coordinates. The displacement of axion from its minimum before the time of the beginning of the oscillation then determines the misalignment angle $\theta\equiv\phi_{\rm os}/\f$. We define the time of the beginning of the oscillation $t_{\rm os}$ as the time when $m=2H_{\rm os}$ \cite{Marsh:2015xka}, where $m$ is the mass of the axion. We then choose the origin of the time coordinate so that $t_{\rm os}=0$ and that the scale factor can be written as $a(t)=(mt+1)^{1/2}$. Note that since the oscillation of the axion is semi-harmonic in  cosmic time, the analysis is more transparent in terms of the cosmic time  compared to the conformal time coordinate. However, the equations of motion for the tensor modes take simpler forms in conformal time which is the topic of the next section.

The quantity of interest  is the number density of the dark photons produced by an oscillating axion in an expanding background. To define the number of particles we need to quantize the gauge field. In this regard, first we define the canonical field $X_i\equiv\sqrt{a}A_i$ which is then expanded in terms of the creation and annihilation operators in momentum space as follows 
\eq{
\label{eq:Xcaop}
X_i(t,\xv)=\sum_{\lambda=\pm}\int\frac{\dd[3]{k}}{(2\pi)^{3/2}}\varepsilon_i^\lambda(\kv)\left(\chi_{k,\lambda}(t) \hat{a}_{\kv,\lambda}+\chi^*_{k,\lambda}(t) \hat{a}^\dagger_{-\kv,\lambda}\right)e^{i\kv.\xv}\,,
}
where $\varepsilon_i^\lambda(\kv)$ are the polarization vectors properties of which are presented in appendix \ref{app:pol}. The effects of time-dependent background are completely encoded in the mode functions $\chi_{k,\lambda}(t)$. Note that due to isotropy the mode functions $\chi_{k,\lambda}$ are functions of the magnitude of the wave vector $k=|\kv|$. Substituting Eq. \eqref{eq:Xcaop} in Eq.~\eqref{eq:Ai}, the mode functions satisfy the equations of motion of the form of a harmonic oscillator with a time-dependent frequency
\eq{
\label{eq:chi}
\ddot{\chi}+\omega_{\rm ph}^2\chi=0\,,\qquad\quad\omega_{\rm ph}^2\equiv\frac{k^2}{a^2}-\lambda\frac{{\alpha}_{a}\dot{\bar{\phi}}}{\f}\frac{k}{a}-\frac{1}{4}(H^2+2\dot{H})\,,
} 
where we have omitted the indices of the mode function(s) for brevity while we keep in mind that the behaviour of the two polarizations are different as the frequency is an explicit function of $\lambda$ which can be $\pm1$ corresponding to right or left-handed photon. The mode function must also satisfy the normalization condition $\chi\dot{\chi}^*-\chi^*\dot{\chi}=i$ to ensure the standard commutation relations for creation and annihilation operators. Furthermore, we assume that photons live in the Bunch-Davies vacuum state annihilated by all $\hat{a}_{\kv,\lambda}$.   

If $\omega_{\rm ph}^2>0$ then we can always write the solution of Eq.~\eqref{eq:chi} for the mode function and its time derivative as \cite{Zeldovich:1971mw,Shtanov:1994ce,Mukohyama:1997bd}
\eqa{
\label{eq:wkbexact}
\chi&=\frac{1}{\sqrt{2\omega_{\rm ph}}}\left(\alpha(t)e^{-i\int\dd{t'}\omega_{\rm ph}}+\beta(t)e^{+i\int\dd{t'}\omega_{\rm ph}}\right)\,,\\
\label{eq:wkbexact2}
\dot{\chi}&=-i\sqrt{\frac{\omega_{\rm ph}}{2}}\left(\alpha(t)e^{-i\int\dd{t'}\omega_{\rm ph}}-\beta(t)e^{+i\int\dd{t'}\omega_{\rm ph}}\right)\,,
}
which imply a set of differential equations for time-dependent coefficients $\alpha(t)$ and $\beta(t)$ as follows
\eq{
\label{eq:alpha-beta}
\dot{\alpha}=\frac{\dot{\omega}_{\rm ph}}{2\omega_{\rm ph}}e^{2i\int\dd{t'}\omega_{\rm ph}}\beta\,,
\qquad\quad\dot{\beta}=\frac{\dot{\omega}_{\rm ph}}{2\omega_{\rm ph}}e^{-2i\int\dd{t'}\omega_{\rm ph}}\alpha\,.
}
The normalization condition of the mode function implies $|\alpha|^2-|\beta|^2=1$. It can be shown that by defining the instantaneous annihilation operator
\eq{
\label{eq:atop}
\hat{\mathfrak{a}}(t) \equiv
\alpha(t)\hat{a}_{\kv,\lambda}+\beta(t)^*\hat{a}^\dagger_{\kv,\lambda}\,,
}
the Hamiltonian can be diagonalized and the number density of photons with definite momentum and helicity at each time is
\eq{\label{n}
n_{k,\lambda}=\bra{0}\hat{\mathfrak{a}}^\dagger(t) \hat{\mathfrak{a}}(t)\ket{0}=|\beta|^2
=\frac{1}{2\omega_{\rm ph}}(|\dot{\chi}|^2+\omega_{\rm ph}^2|\chi|^2)-\frac{1}{2}\,.
}
We assume that the initial state has no particle which means $\beta(0)=0$ and $\alpha(0)=1$.
As we shall see shortly later, $\omega_{\rm ph}^2$ can also become negative for some time intervals. In that case, we define $\Omega_{\rm ph}^2=-\omega_{\rm ph}^2>0$ by means of which we can express the mode function as
\eqa{
\label{eq:exacttach}
\chi&=\frac{1}{\sqrt{2\Omega_{\rm ph}}}\left({\rm a}(t)e^{-\int\dd{t'}
\Omega_{\rm ph}}+{\rm b}(t)e^{\int\dd{t'}\Omega_{\rm ph}}\right) \,,\\
\dot{\chi}&=-\sqrt{\frac{\Omega_{\rm ph}}{2}}\left({\rm a}(t)e^{-\int\dd{t'}\Omega_{\rm ph}}-{\rm b}(t)e^{\int\dd{t'}\Omega_{\rm ph}}\right)\,,
}
where the time-dependent coefficients ${\rm a}(t)$ and ${\rm b}(t)$ satisfy
\eq{
\dot{{\rm a}}=\frac{\dot{\Omega}_{\rm ph}}{2\Omega_{\rm ph}}e^{2\int\dd{t'}\Omega_{\rm ph}}{\rm b}(t)\,,\qquad\quad\dot{{\rm b}}
=\frac{\dot{\Omega}_{\rm ph}}{2\Omega_{\rm ph}}e^{-2\int\dd{t'}\Omega_{\rm ph}}{\rm a}(t)\,,
}
with the constraint ${\rm a}{\rm b}^*-{\rm a}^*{\rm b}=i$. We do not extend the notion of particle number to the regime where $\omega_{\rm ph}^2<0$ which does not seem to be well defined \cite{Rajeev:2017uwk} and only measure the number density in the regime when $\omega_{\rm ph}^2>0$.

\subsection{Solving the mode function}\label{subsec-mode-function}
In order to obtain the number density of the produced dark photons, we need to solve the equations of motion for the mode function Eq.~(\ref{eq:chi}) with the initial conditions of  zero particle number density. A solution in the form of Eq.~\eqref{eq:wkbexact} for $\omega_{\rm ph}^2>0$ or Eq.~\eqref{eq:exacttach} for $\omega_{\rm ph}^2<0$, despite being exact, may not be very useful since solving equations for the coefficients $\alpha$, $\beta$ and ${\rm a}$, ${\rm b}$ are as difficult as the equation for the mode function itself. In this subsection we explore the general properties of Eq.~\eqref{eq:chi} which is characterized by the effective photon frequency $\omega_{\rm ph}^2$ and then find approximate solutions based on the analytic method employed in preheating scenarios after inflation in the next subsection \cite{Kofman:1994rk,Kofman:1997yn}.

Let us rewrite Eq.~\eqref{eq:chi} in a more appropriate form by substituting from \eqref{axion-osclliation} and defining the dimensionless variable $x\equiv mt$ as
\eq{
	\label{eq:chiy}
	\dv[2]{\chi}{x}+\omega^2\chi=0\,,
} 
where we have defined the dimensionless frequency
\begin{eqnarray}
\label{eq:omega2new}
\omega^2 \equiv \Big(\frac{\omega_{\rm ph}}{m}\Big)^2 = \frac{\kappa^2}{a^2}+\lambda\frac{\kappa\vp}{a^{5/2}}
\sin{x} - \frac{H^2+2\dot{H}}{4 m^2} \approx 
\frac{\kappa^2}{a^2}+\lambda\frac{\kappa\vp}{a^{5/2}}\sin x\,,
\end{eqnarray} 
in terms of the dimensionless parameters $\kappa\equiv k/m$ and $\vp\equiv {\alpha}_{a}\phi_{\rm os}/\f$. In deriving the above result, we have neglected the term suppressed by $H/m$ in computing $\dot{\bar{\phi}}$ from Eq.~\eqref{axion-osclliation}. Further, in the last step in Eq.~\eqref{eq:omega2new} we have also neglected the mass term induced by the expansion of the universe. The justification is as following: in RD universe we have $\dot{H}\sim H^2$ and therefore the mass term induced by the expansion is proportional to $H^2/m^2$ which is less than unity after the axion field started its oscillation and evolves like $a^{-4}$ which falls off  faster than the other terms. Moreover, we will show that the interesting range of parameters are $\kappa\gtrsim1$ and $\vp\gtrsim1$. As a result, we can safely neglect the mass term induced by the expansion of the universe in our upcoming analytic investigations. Since the Hubble parameter is smaller than the mass scale it is reasonable as a first approximation to set $a=1$ and neglect the effects of expansion. We then add the effects of the expansion in an appropriate manner. In this approximation we can write Eq.~\eqref{eq:chiy} in the standard form of the so-called Mathieu equation \cite{McLachlan:1964}
\eq{
\label{eq:matheiu}
\dv[2]{\chi}{z}+(A_{\rm m}-2q_{\rm m}\cos(2z))\chi=0\,,
} 
where we have defined $z\equiv \frac{1}{2}x-\frac{\pi}{4}$, $A_{\rm m}\equiv 4\kappa^2$ and $q_{\rm m}\equiv2\kappa\vp$. In deriving the above equation we have set $\lambda=-1$. The equation for the other polarization is obtained by the transformation $q_{\rm m}\to-q_{\rm m}$, but this does not change characteristics of Mathieu equation since it can be realized by a shift in time $z\to z+\pi/2$. Hence, if we neglect the expansion and also the backreaction of photons to the dynamics of $\bar{\phi}$ (right hand side of Eq.~\eqref{eq:phibar}), then there is \emph{no parity violation}. We will see that this is no longer true if we take into account the expansion of the universe \cite{Sorbo:2011rz,Adshead:2015pva} and the backreaction.

The interesting feature of Eq.~\eqref{eq:matheiu} is that although its coefficient is periodic with period equal to $\pi$, the solution is not necessarily periodic but in general has the form $e^{\mu z}p(z)$ where $p(z)$ is a periodic function and $\mu$ is a complex number called the characteristic exponent (see appendix~\ref{app:floquet} for more details). If $\mu$ happens to have a positive real part then the amplitude of $\chi$ grows exponentially and the solution is called unstable while if $\mu$ is purely imaginary then $\chi$ is bounded and stable. The boundary, in the parameter space, between stable and unstable solutions can be found by seeking periodic solutions of Eq.~\eqref{eq:matheiu} with period equal to $\pi$ or $2\pi$ (see appendix~\ref{app:floquet}). Fig.~\ref{fig:chart} shows stable regions as shaded area for different values of $\vp$ and $\kappa$ while the white area corresponds to unstable solutions. In the case of instability, the number of particles grows exponentially like $e^{2\mu z}$. This phenomena is usually called parametric resonance. The process of exponential particle production continues until backreaction of photons becomes important in the dynamics of the oscillating field $\bar{\phi}(t)$. For very small $\vp$, unstable solutions are obtained for a narrow band around $\kappa=n/2$ for $n=1,2,\dots$ which is called narrow resonance while for larger $\vp$ the instability bands include a wider range of $\kappa$ thus called broad resonance.     
\fg{
\centering
\includegraphics[width=0.43	\textwidth]{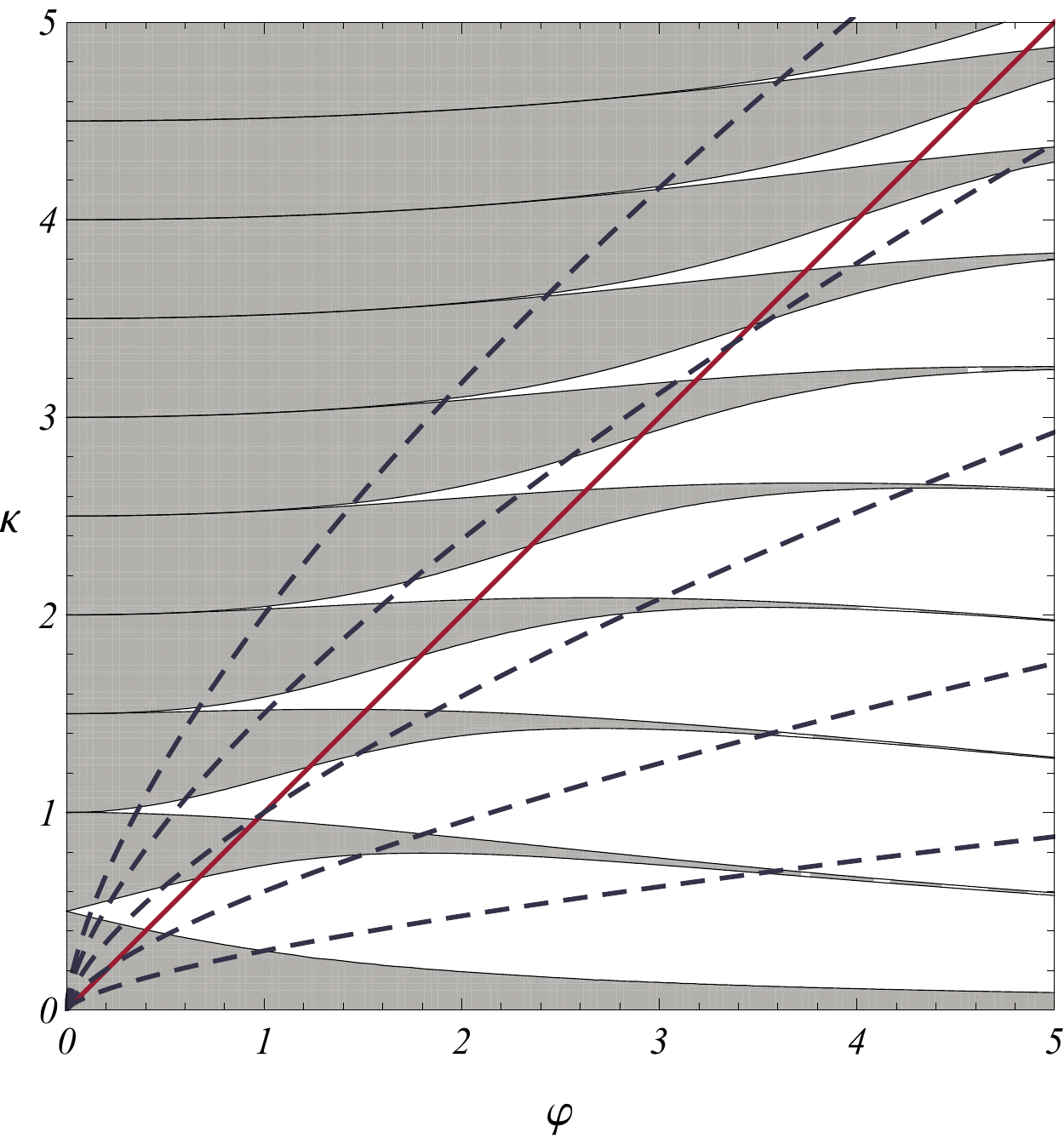}
\caption{\small Stability chart of Eq.~\eqref{eq:matheiu} in terms of $\kappa\equiv k/m$ and $\vp\equiv {\alpha}_{a}\phi_{\rm os}/\f$. Shaded area shows the stable regions while white regions correspond to unstable solutions. The thick red line represents $\kappa=\vp$. Below this line $\omega^2$ becomes negative in each period. Dashed blue curves show adiabatic evolution due to the background expansion. As time goes on, any point approaches the origin through these curves.}
\label{fig:chart}
}
The nature of instabilities are different in narrow and broad resonance regimes \cite{Kofman:1997yn,Kofman:1996mv}. In the narrow resonance band  particles are produced gradually all the time while in the latter case they are produced explosively in a short interval of time in each period of oscillation. The thick red line in Fig.~\ref{fig:chart} corresponds to $\kappa=\vp$. Below this line $\omega^2$ becomes negative for a time interval in each period of oscillation while above this line $\omega^2$ is always positive in each period. The general expectation is that when $\omega^2<0$ the solutions grow exponentially and particle production can be very efficient. This is consistent with Fig.~\ref{fig:chart} where most of the instability bands are below the line $\kappa=\vp$. However, even below this line there are narrow stable regions which occur due to the destructive interference between positive and negative frequency solutions during the time $\omega^2>0$ \cite{Dufaux:2006ee}.

Parametric resonance becomes more and more inefficient in the course of time if we take into account the expansion of the universe. This can be studied adiabatically by considering a time dependence in the parameters of the form
\eq{
\vp(t)=\frac{\vp}{a^{3/2}}\,,\qquad\kappa(t)=\frac{\kappa}{a}\,,
}       
which are related to each other as $\kappa(t)\propto\vp(t)^{2/3}$ where the constant of proportionality is fixed by initial conditions. The dashed blue lines in Fig.~\ref{fig:chart} show this time evolution. Even if a point $(\vp(t),\kappa(t))$ is initially located in an unstable region, the background expansion flows this point along the curves $\kappa(t)\propto\vp(t)^{2/3}$ through many stability/instability bands and finally to the region where there is no particle production. In particular, if $\vp>\kappa$ so that the point is initially ($x=0$) located below the thick black line in Fig.~\ref{fig:chart}, after the time $x_{\rm p}$ given by 
\eq{
\label{eq:apositive}
a_{\rm p}=(\vp/\kappa)^2\,,
}
it crosses the red line in Fig.~\ref{fig:chart} which means that $\omega^2$  becomes always positive. For a fixed $\vp$, different momenta corresponding to different $\kappa$ have different fates. If $\kappa$ is very large (but still less than $\vp$) then $x_{\rm p}$ is very small and the mode crosses the line $\kappa=\vp$ very soon and spends most of its time in the stable shaded area of Fig.~\ref{fig:chart}. Its last chance for particle production is  a short period in the first instability band, i.e. the one opening up at around $\kappa=1/2$. As a result, we expect that $\kappa$ with the most efficient particle production is the one that leaves the first band at a time around $x_{\rm p}$ which means that it has spent all its particle production history below the line $\kappa=\vp$. Since this usually happens in the last stages of the parametric resonance, we expect that the resonance band is narrow where Eq.~\eqref{eq:band1} is valid for the width of the first band. Then particle production terminates at $x=x_{\rm en}$ when the lower bound of the inequality in Eq.~\eqref{eq:band1} is reached with
\eq{
\label{eq:aen}
\frac{4\kappa^2}{a_{\rm en}^2}=1-\frac{2\kappa\vp}{a_{\rm en}^{5/2}}\,.
}    
If we demand $a_{\rm en}=a_{\rm p}$ then the corresponding momentum $\kappa_*$ is given by
\begin{eBox}
\eq{
\label{eq:k*}
\kappa_*=\frac{\vp^{2/3}}{6^{1/6}}\sim(\vp/1.5)^{2/3}\,.
}
\end{eBox}
Since the amplitude of the unstable mode function is of exponential form we expect that the number of photons corresponding to other momenta be exponentially suppressed compared to $\kappa_*$. As a result, the spectrum of produced photons has a peak at $\kappa_*$. Another way to obtain a similar result is an intuitive reasoning due to \cite{Machado:2018nqk} which we review here for the completeness. At each time the range of momenta which are below the thick red line of Fig.~\ref{fig:chart}, is the range $0<\kappa<\vp/\sqrt{a}$. The growth rate is intuitively controlled by the amplitude of $|\omega|$ which is maximized in the middle of the above range at $\vp/(2\sqrt{a})$ with the maximum value $|\omega_{\rm max}|=\vp/(2a^{3/2})$. On the other hand, the time scale for one oscillation is $m^{-1}$ which means that in order to have efficient particle production we need $|\omega|\gtrsim{1}$ (or equivalently $|\omega_{\rm ph}|\gtrsim{m}$). As a result, we find that the efficient particle production ends when the scale factor approaches $(\vp/2)^{2/3}$. The specific momentum which maximizes $|\omega|$ at this time is $(\vp/2)^{2/3} \sim \kappa_*$ which is consistent with Eq.~\eqref{eq:k*}. As we will see in more detail in the following section, for the tachyonic regime we will have $\kappa_*\gtrsim1$. This seems to be counter-intuitive since from perturbation theory we expect the axion particle with mass $m$ can decay into two photons with momentum $m/2$ corresponding to $\kappa=1/2$. However, parametric resonance is a nonperturbative effect and the naive intuition from the perturbative interactions no longer works here. Even in the narrow resonance regime where parameters might fit into the perturbative analysis, the Bose factor due to the large number density of axion particles enhances the rate of the particle production \cite{Kofman:1997yn}.         

Apart from the non-perturbative particle production that we discussed in this subsection,  in principle, perturbative decay of the axion to the dark photons can also happen similarly to what happens for inflaton during perturbative reheating after non-perturbative preheating \cite{Kofman:1994rk,Kofman:1997yn}. In appendix \ref{app:perturbative-decay}, based on the analytic formalism that we extended in this section, we compute the decay rate of the axion to dark photons. Using the decay rate, we consider the possibility for the perturbative decay of the axion. As it is well known, for the favourable ranges of the axion mass and coupling, the perturbative decay may not happen in the time scales shorter than the age of the universe (see appendix \ref{app:perturbative-decay}). In this regard, axion remnants can serve as part of the dark matter \cite{Agrawal:2017eqm,Agrawal:2018vin,Machado:2018nqk,Marsh:2015xka}.

In the next subsection, we find semi-analytic solution of \eqref{eq:chi} for the mode function. The readers who are interested only in the results for the GWs may skip the next subsection and directly move to subsection \ref{subsec-backreaction}. 

\subsection{Analytic computation of number density}\label{vector-field-mode-function}
Although the picture presented in the previous subsection gives a correct qualitative description, it is not very useful in obtaining the analytic solutions for the equations of motion for the mode functions \eqref{eq:chi}. In this subsection, following the method of successive scattering introduced in Refs.~\cite{Kofman:1997yn,Dufaux:2006ee}, we find analytic solutions for the mode functions of the dark photons. The starting point is to consider the exact forms \eqref{eq:wkbexact} and \eqref{eq:exacttach} which serve as natural basis for the adiabatic approximation. It is easy to see that in the domain of validity of the adiabatic approximation, the coefficients $\alpha(t)$, $\beta(t)$, ${\rm a}(t)$ and ${\rm b}(t)$ are approximately constant in time and thus no particle production occurs. As a result, it is sufficient to seek regions where the adiabatic approximation breaks down  and then compute the change in the particle number in each event. A caveat is that this procedure does not work  in the narrow resonance regime where the adiabatic approximation almost always holds (see Ref.~\cite{Kofman:1997yn}). However,  since in most part of the analysis we are interested in the broad resonance regime, this is enough for our purposes. 

In the broad resonance regime, the adiabatic approximation holds for most of the time evolution of the mode function $\chi$. As a result, we can approximately write the solutions for \eqref{eq:chi} as
\eq{
\label{eq:posmass}
\chi\approx\frac{\alpha}{\sqrt{2\omega(x)}}\exp(-i\int^x\dd{x'}\omega(x'))+
\frac{\beta}{\sqrt{2\omega(x)}}\exp(+i\int^x\dd{x'}\omega(x'))\,,
} 
during the time in which $\omega^2>0$. Note that the coefficients $\alpha$ and $\beta$ are constant in adiabatic approximation. Comparing the above form  with Eq.~\eqref{eq:wkbexact} we find that as long as the adiabatic approximation is valid the number density $n=|\beta|^2$ does not change. The integration in the exponential is only taken over times when $\omega^2>0$. As we have seen before, taking $\kappa<\vp$ then $\omega^2$ becomes negative for a time interval in each oscillation of $\bar{\phi}$. If the adiabatic approximation \emph{holds} during that interval, the solution would take the same form as Eq.~\eqref{eq:exacttach} and we have
\eq{
\label{eq:negmass}
\chi\approx\frac{{\rm a}}{\sqrt{2\Omega(x)}}\exp(-\int^x\dd{x'}\Omega(x'))+
\frac{{\rm b}}{\sqrt{2\Omega(x)}}\exp(+\int^x\dd{x'}\Omega(x'))\,,
} 
where $\Omega^2=-\omega^2>0$ and the integration is only taken over the period where $\omega^2$ is negative. We call the regime, in which the approximate solution of the form Eq.~\eqref{eq:negmass} holds, \emph{tachyonic instability}. As we already discussed, during the interval $\omega^2<0$ the particle number is not well-defined but the mode function grows exponentially and results in typically large amount of new particles after $\omega^2$ becomes positive once again. 

Approximate solutions of the form \eqref{eq:posmass} or \eqref{eq:negmass} are valid for the regions $\omega^2>0$ and $\omega^2<0$ respectively and needed to be sewn together in regions where adiabatic approximation breaks down, i.e.
\eq{
	\left\vert\frac{1}{\omega^2}\frac{d\omega}{dx}\right\vert\gtrsim1\,,\qquad \mbox{or/and} \qquad
	\left\vert\frac{1}{\omega^3} \frac{d^2\omega}{dx^2} \right\vert \gtrsim1\,.
} 
\begin{figure}
\centering
\includegraphics[width=0.7\textwidth]{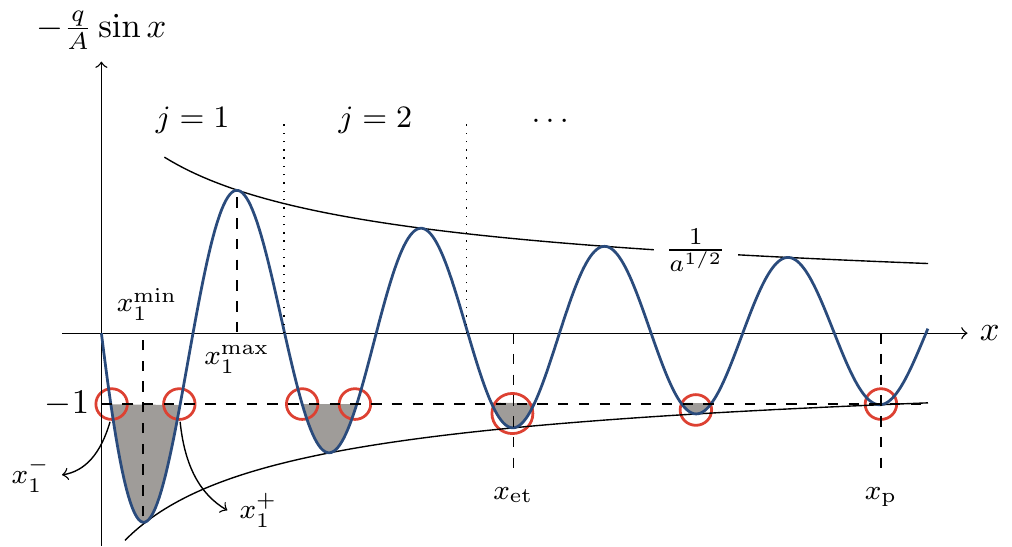}
\caption{\small The effective frequency of the dark photon is of the form $\omega^2=A\left(1+\lambda\frac{q}{A}\sin x\right)$. Time variation of the oscillating term $\lambda\frac{q}{A}\sin x$ is shown above for $\lambda=-1$.
Below the horizontal dashed line, the effective frequency of photon $\omega^2$ becomes negative and is shown with the shaded areas. Regions where adiabatic approximation breaks down are marked with red circles. The end of tachyonic regime is specified by $x_{\rm et}$ (see Eq.~\eqref{eq:adappmin} and the discussion below it) and after $x_{\rm p}$ (see Eq.~\eqref{eq:apositive}) the effective frequency $\omega^2$ will be always positive.}
\label{fig:delta}
\end{figure}
Let us write $\omega^2$ in Eq.~\eqref{eq:omega2new} as  
\eq{
\label{eq:omega2re}
\omega^2=A\left(1+\lambda\frac{q}{A}\sin x\right)\,,\qquad\text{with}\qquad A=\frac{\kappa^2}{a^2}\,,\quad q=\frac{\kappa\vp}{a^{5/2}}\,,
}
where $A$ and $q$ defined here are different from $A_{\rm m}$ and $q_{\rm m}$ defined under Eq.~\eqref{eq:matheiu} for the standard Mathieu equation. The second term in the parenthesis is shown in Fig.~\ref{fig:delta} for $\lambda=-1$ which is only a sine function with decreasing amplitude $\frac{q}{A}=\frac{\vp}{\kappa\sqrt{a}}$ as the universe expands. The $j-$th period of $\bar{\phi}$ oscillation is during the interval $(x_{j-1},x_j)$ with $x_j=2\pi j$. If we assume that $\frac{q}{A}>1$ then as shown by shaded area in Fig.~\ref{fig:delta}, $\omega^2$ becomes negative in $j-$th oscillation during the times between $x_j^-$ and $x_j^+$ where
\eq{
\label{eq:xjpm}
x_j^-=\overline{x}_j+2\pi(j-1)\,,\qquad x_j^+=\pi-\overline{x}_j+2\pi(j-1)\,,\quad\text{with}\quad \overline{x}_j=\arcsin\frac{A_j}{q_j}\,,
}
in which we have approximately set $A$ and $q$ to their values at $x_j^{\rm min}$ denoted respectively by $A_j$ and $q_j$ \cite{Abolhasani:2009nb}. Here, $x_j^{\rm min}\equiv 2\pi j+(\lambda-2)\pi/2$ corresponds to the minimum of $\lambda\sin x$ in the $j-$th period. It is then evident that the adiabatic approximation breaks down around the points $x_j^{\pm}$ where $\omega^2\approx0$ which are specified by the red circles in Fig.~\ref{fig:delta}. Thus we can use expressions \eqref{eq:posmass} and \eqref{eq:negmass} respectively for positive and negative $\omega^2$ except around the turning points $x_j^\pm$. Around the turning points, the approximate solution for the mode function can be found in terms of Airy functions $Ai(x)$ and $Bi(x)$ by Taylor expanding $\omega^2$ up to linear order.  We can use the asymptotic form of Airy functions to match the coefficients of Eqs.~\eqref{eq:posmass} and \eqref{eq:negmass}. The procedure is very similar to connecting adiabatic solutions for the wave function in classically allowed and forbidden regions in quantum mechanics. An equivalent approach would be to extend the definition of $\chi$ to the complex $x-$plane \cite{Dufaux:2006ee,landau:qmech}. By repeating this procedure we can   connect the coefficients of Eq.~\eqref{eq:posmass} during $j-$th oscillation $(\alpha_j,\beta_j)$ to the coefficients of $j+1-$th oscillation $(\alpha_{j+1},\beta_{j+1})$ as
\cite{Dufaux:2006ee}
\eq{
\label{eq:mattach}
\begin{bmatrix}
\alpha_{j+1}\\\beta_{j+1}
\end{bmatrix}
=e^{X_j}
\begin{bmatrix}
1&ie^{2i\theta_j}\\
-ie^{-2i\theta_j}&1
\end{bmatrix}
\begin{bmatrix}
\alpha_{j}\\\beta_{j}
\end{bmatrix}\,,
}       
where
\eq{
\label{eq:Xjdef}
X_j\equiv\int_{x_j^-}^{x_j^+}\dd{x'}\Omega(x')\,,
}
and $\theta_j \equiv \theta_0+\sum_j\Theta_j$ is the accumulated phase during the time intervals of positive $\omega^2$ up to the point $x_j^-$ in which $\theta_0$ is some initial phase and for $j\geq1$
\eq{
\label{eq:thetaj}
\Theta_j \equiv \int_{x_{j}^+}^{x_{j+1}^-}\dd{x'}\omega(x')\,,
}
is the accumulated phase during $j-$th oscillation. 

If we set $\alpha_0=1$ and $\beta_0=0$ initially, corresponding to the initial vacuum state, then in $j-$th oscillation the number density of particles is obtained to be \cite{Abolhasani:2009nb}
\eq{
\label{eq:nj}
n_j=|\beta_{j}|^2=\exp(2\sum_{\ell=1}^{j}X_\ell)\prod_{\ell=1}^{j-1}(2\cos\Theta_\ell)^2\,.
}
\fg{
\centering
\includegraphics[width=0.63\textwidth]{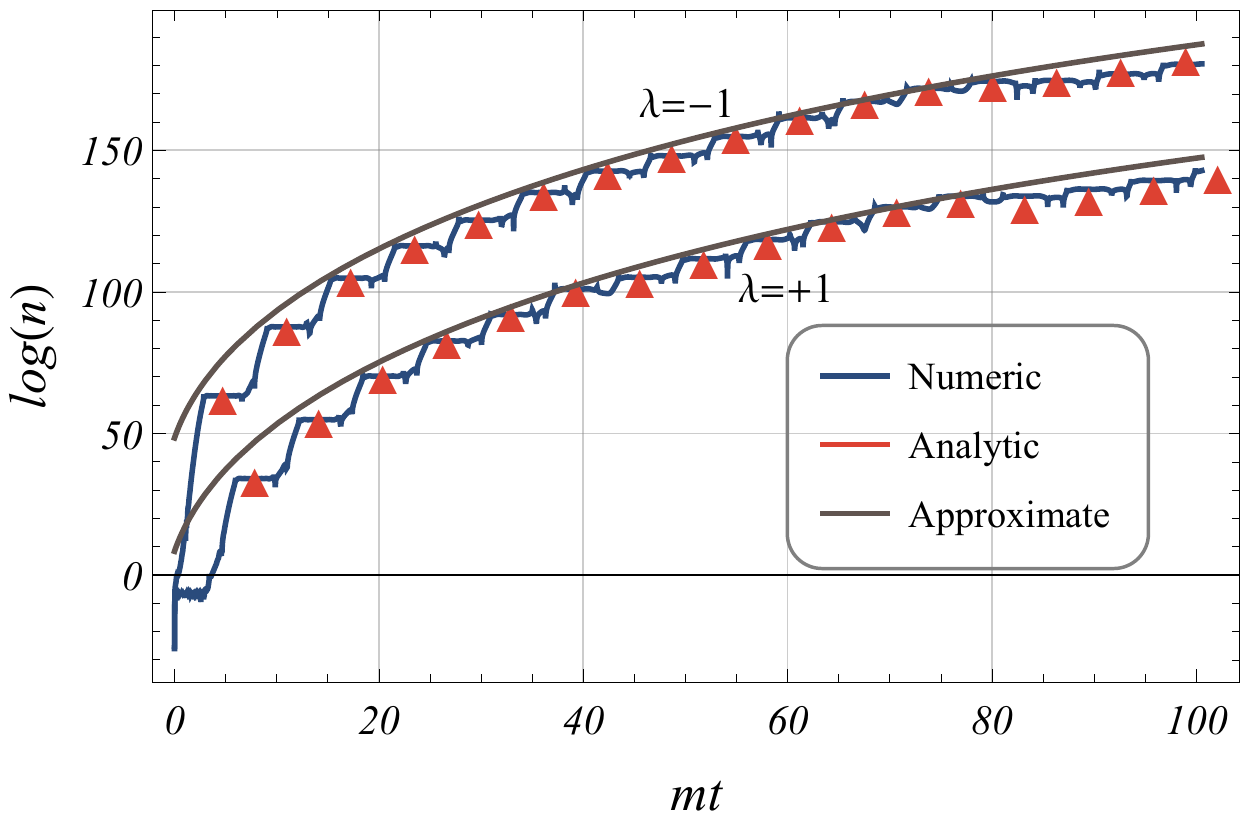}
\caption{\small Logarithm of the dark photons' number density during the tachyonic regime as a function of time for photons (both polarizations) with wave-number $k_*=m\kappa_*$ given by Eq.~\eqref{eq:k*}. This is expected to be the maximum of the dark photon spectrum. For this plot we have used $\vp=70$. The blue curve is obtained by solving Eq.~\eqref{eq:chiy} numerically. The Red triangles show the result of successive scattering matrix method in Eq.~\eqref{eq:nj}. The gray curve is the approximate formula of Eq.~\eqref{eq:sumX}.}
\label{fig:nj}
}
This can be used to compute the number density of dark photons as a function of time by computing the integrals \eqref{eq:Xjdef} and \eqref{eq:thetaj}.  Fig.~\ref{fig:nj} shows the number density of dark photons for both polarizations obtained from numerical calculation of the mode function (blue curve). The red triangles show the number density obtained from successive multiplication of matrices in Eq.~\eqref{eq:mattach} (or Eq.~\eqref{eq:nj}) which is in very good agreement with the numerical result. As is evident from Fig.~\ref{fig:nj}, left-handed polarization corresponding to $\lambda=-1$ dominates over the right-handed one. The reason is that if we assume $\vp>0$, then the point $x_j^{\rm min}$ happens earlier for the left-handed polarization ($\lambda=-1$) compared to the right-handed one ($\lambda=+1$) by an amount $\Delta x=\pi$ in each oscillation (see Fig.~\ref{fig:delta}). As a result, the magnitude of $X_j$ is relatively larger for the left-handed polarization during the period of exponential growth and the total number of particles is exponentially larger. Needless to say  the situation is reversed if we choose $\vp$ to be negative\footnote{Note that even without expansion we have parity violation if we consider the backreaction of the produced photons  on the dynamics of $\bar{\phi}$. In this case the amplitude of $\bar{\phi}$ decreases because of the energy transfer to the dark photons and the polarization which becomes tachyonic dominates sooner again.}. 

The number density in the $j-$th oscillation is mostly controlled by the exponential factor in Eq.~\eqref{eq:nj} so in the following we try to find a closed form expression for it. First, note that from Eq.~\eqref{eq:Xjdef} we can compute 
\eq{
\label{eq:Xj}
X_j=4\sqrt{q_j-A_j}E\left(\frac{\pi}{4}-\frac{1}{2}\overline{x}_j;\frac{2q_j}{q_j-A_j}\right)
\approx2.4\sqrt{q_j}\left(1-\frac{A_j}{q_j}\right)\,,
}
where $E(\phi;m)$ is the incomplete elliptic function of the second kind and we have used the approximation introduced in Ref.~\cite{Dufaux:2006ee}. As mentioned above, $A$ and $q$ are approximately evaluated at $x=x_j^{\rm min}$ for which the scale factor is $a_j=\left(2\pi j+(\lambda-2)\pi/2+1\right)^{1/2}$ for both polarizations $\lambda=\pm1$. Also for $\lambda=-1$, the scale factor is evaluated at $\Delta x=\pi$ earlier in each period. Indeed $a_j$ must also carry an index for polarization but to avoid involved notation we do not show that explicitly. After substitution of $A_j$ and $q_j$ from Eq.~\eqref{eq:omega2re} in terms of $a_j$, the summation in the exponential of Eq.~\eqref{eq:nj} can be computed by using the definition of the Hurwitz-Zeta function $\zeta_s(r)=\sum_{j=0}^{\infty}1/(j+r)^s$ \cite{atlas:2000}. After a couple of lines of algebra we obtain
\begin{eBox}
\eq{
\label{eq:sumX}
\sum_{\ell=1}^{j}X_\ell=2.4\frac{\sqrt{\kappa\vp}}{(2\pi)^{5/8}}\left[\zeta_{5/8}\big(r_\lambda\big)-\zeta_{5/8}\Big(\frac{a_j^2}{2\pi}+1\Big)-(2\pi)^{1/4}\frac{\kappa}{\vp}\Big(\zeta_{3/8}(r_\lambda)-\zeta_{3/8}\Big(\frac{a_j^2}{2\pi}+1\Big)\Big)\right]\,,
}    
\end{eBox}
where we have defined $r_\lambda \equiv(2+(2+\lambda)\pi)/4\pi$. In the final expression we have replaced the total number of oscillations, $j$, in terms of $a_j$ in the argument of the Hurwitz-Zeta function. 

Since the process of production of dark photons is very short it is useful to find approximate expression for \eqref{eq:sumX} for a few number of oscillations. In this case, we can obtain approximate expression for the Hurwitz-Zeta functions as
\eqa{
\zeta_{5/8}(a^2/2\pi+1)&\approx-1.65-0.6a \,, \\
\zeta_{3/8}(a^2/2\pi+1)&\approx-0.8-0.41a-0.045a^2 \,,
}
for the range $1\lesssim a\lesssim10$. As a result, we can approximately write Eq.~\eqref{eq:sumX} as
\eq{
\label{eq:sumXapp}
\sum_{\ell=1}^{j}X_\ell\approx\sqrt{\kappa\vp}\left[c_\lambda+\tilde{c}_\lambda\frac{\kappa}{\vp}+\left(0.45-0.49\frac{\kappa}{\vp}\right)a_j-0.05\frac{\kappa}{\vp}a_j^2\right]\,,
}
where the polarization dependent coefficients are $c_\lambda=-0.425\lambda+0.175$ and $\tilde{c}_\lambda=0.5\lambda-0.32$ for $\lambda=\pm1$. The approximate number density obtained from Eq.~\eqref{eq:sumXapp} by neglecting the factors containing phases in Eq.~\eqref{eq:nj} is shown in Fig.~\ref{fig:nj} by the gray curve which is in reasonable agreement with the numerical result. Note that by looking at Eqs.~\eqref{eq:sumX} or \eqref{eq:sumXapp} in the exponent appearing in Eq.~\eqref{eq:nj}, we can spot a dependency of the form $\exp(-1/\sqrt{\vp})$ on the coupling between dark photon and the axion. This confirms our earlier claim that the particle production happens in the non-perturbative regime \cite{Kofman:1997yn}.

Having obtained the approximate number density from Eq.~\eqref{eq:sumXapp} we can estimate the amount of parity violation by comparing the exponent for two polarizations. Note that if we compare the values of Eq.~\eqref{eq:sumXapp} after a couple of oscillations, the dependency of $a_j$ on $\lambda$ becomes unimportant so we can write
\eq{
\frac{A_+}{A_-}\sim\exp(\sqrt{\kappa\vp}\left[-0.85+\frac{\kappa}{\vp}\right])\,.
} 
Note that, as discussed in the last section, we must have $\kappa<\vp/\sqrt{a}$ for the tachyonic instability to occur, so the above expression is always negative in this regime confirming the subdominance of right-handed polarization compared to the left-handed. Evaluating this for $\kappa_*$ reveals that the amount of parity violation becomes stronger for larger $\vp$ as expected. 

As we include more and more oscillations, the amplitude of the sine term in Eq.~\eqref{eq:omega2re} decreases and the time interval in which $\omega^2$ is negative shrinks until finally after the time $x_{\rm p}$ computed in Eq.~\eqref{eq:apositive}, $\omega^2$ would be always positive and tachyonic instability disappears. However, as shown schematically in Fig.~\ref{fig:delta}, even before $x_{\rm p}$ is reached the time interval between $x_j^-$ and $x_j^+$ becomes so tiny that the adiabatic approximation breaks down even at the point $x_j^{\rm min}$ \cite{Dufaux:2006ee}. To check this, we compute $\left\vert\frac{1}{\omega^3} \frac{d^2\omega}{dx^2} \right\vert$ at $x_j^{\rm min}$ which reveals that the adiabatic approximation is valid as long as
\eq{
\label{eq:adappmin}
\varsigma^2\equiv\frac{|q_j-A_j|}{\sqrt{q_j/2}}\gtrsim1\,.
}
However, due to the expansion the inequality is soon violated which means that the adiabatic approximation is no longer true around the point $x_j^{\rm min}$. Thus, although $\omega^2$ is still negative around $x_j^{\rm min}$, Eq.~\eqref{eq:negmass} can no longer be exploited. We denote the time when this happens by $x_{\rm et}$ and the corresponding scale factor by $a_{\rm et}$ since this specifies the end of the tachyonic instability regime. For $\kappa_*$ given in Eq.~\eqref{eq:k*} using the condition of Eq.~\eqref{eq:adappmin} we find that $a_{\rm et}\approx0.6\vp^{2/3}$. This is approximately the time when, according to the intuitive description of Ref.~\cite{Machado:2018nqk}, the tachyonic band closes, although as mentioned before, $\omega^2$ is still negative in each period hence	 we do not use this terminology here.\footnote{If following Ref.~\cite{Machado:2018nqk} one sets $\kappa_*=(\vp/2)^{2/3}$ then from Eq.~\eqref{eq:adappmin} we find $a_{\rm et}=(\vp/2)^{2/3}$. This is the same value for the scale factor as what Ref.~\cite{Machado:2018nqk} calls closure of the tachyonic band.}  

\subsubsection*{Semi-tachyonic regime}
After the end of tachyonic instability regime at $x_{\rm et}$, we can still have particle production due to the parametric resonance if parameters happen to be located in one of the instability bands. Since adiabatic approximation is not valid for the whole interval between $x_j^-$ and $x_j^+$ we cannot use Eq.~\eqref{eq:negmass} there. However, in this regime $x_j^+-x_j^-$ is a very small fraction of a period (see Eq.~\eqref{eq:xjpm} and also shaded regions after $x_{\rm et}$ in Fig. \ref{fig:delta}) such that we can instead solve the mode function approximately around the point $x_j^{\rm min}$ and then connect the obtained solution to adiabatic solution of Eq.~\eqref{eq:posmass} far from the point $x_j^{\rm min}$. The sine function can be approximated as a parabola near the point $x_j^{\rm min}$ and the solution of the mode function can be written in terms of the parabolic cylinder function (see appendix \ref{semitach}). The asymptotic form of the mode function is then matched with Eq.~\eqref{eq:posmass}. A very similar problem is addressed in Ref.~\cite{Kofman:1997yn} for the broad parametric resonance regime. The difference is that here $\omega^2$ is negative around the point of approximation, i.e. at $x_j^{\rm min}$ while in Ref.~\cite{Kofman:1997yn} the value of $\omega^2$ is always positive. The details of the calculations are given in appendix \ref{semitach}.

\fg{
	\centering
	\includegraphics[width=0.63\textwidth]{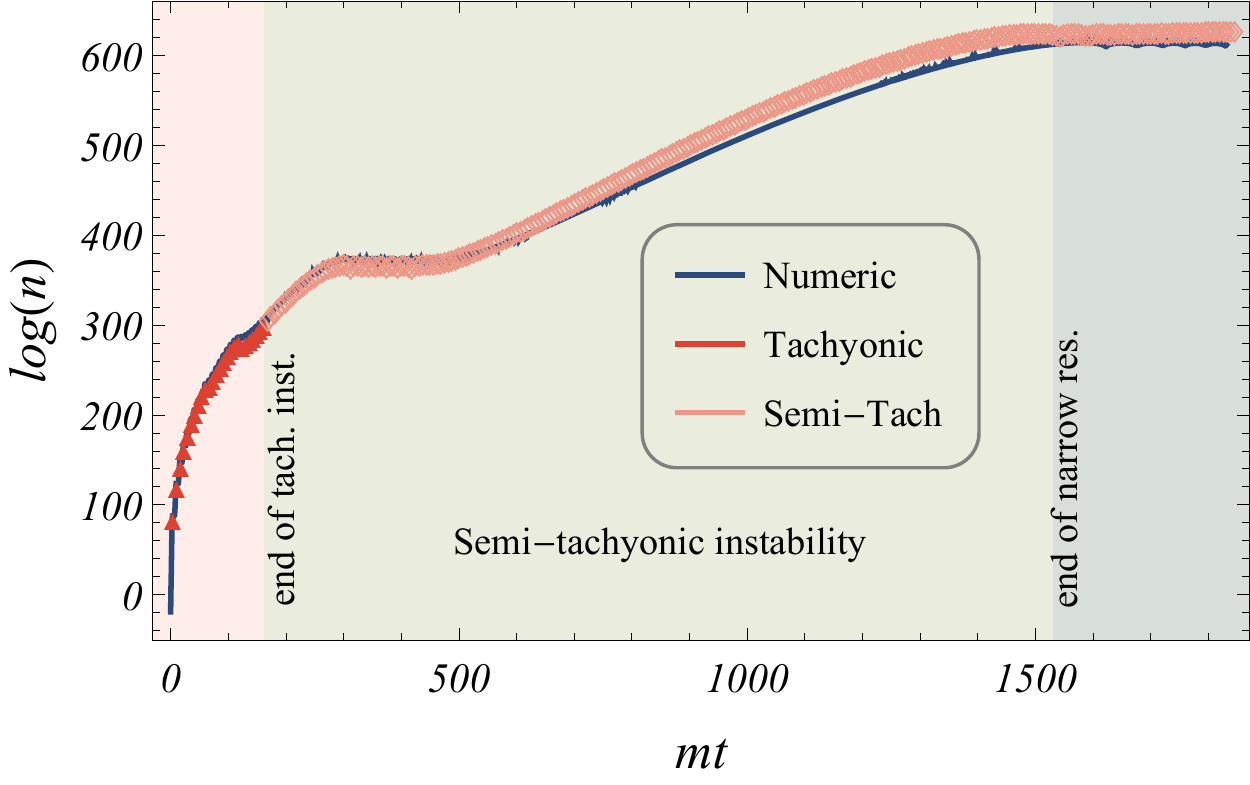}
	\caption{\small Logarithm of the number density of photons for a larger period of time until the end of particle production (ignoring backreaction effects). The photon wave-number is set to $k=m\kappa_*$ and the we have shown only the dominant polarization. For this plot we have set $\vp=100$. The particle production history can be divided into three regimes. Initially, we have tachyonic instability until the time $x_{et}$ (see Eq.~\eqref{eq:adappmin} and below it). Then, we have semi-tachyonic regime until the time $x_{en}$ (see Eq.~\eqref{eq:aen}). After that we do not have efficient particle production. The blue curve is obtained from numeric solution while red and pink point are obtained by the analytic results of Eqs.~\eqref{eq:nj} and \eqref{eq:napp} for the tachyonic and semi-tachyonic regime.}
	\label{fig:njall}
}

From the transfer matrix in Eq.~\eqref{eq:tmat} we can relate $n_{j+1}$ to $n_{j}$, which 
for $n_j\gg1$,  yields 
\eq{
\label{eq:napp}
n_{j+1}=\left[1+2e^{\pi\vrk^2}-2e^{\pi\vrk^2/2}\sqrt{1+e^{\pi\vrk^2}}\sin\theta_{\rm tot}\right]n_j\,,
}
where $\theta_{\rm tot}\equiv \vartheta+2\theta_j+\arg\beta_{j}-\arg\alpha_{j}$ in which $\vartheta$ and $\theta_j$ are defined in Eq.~\eqref{eq:vartheta} and below Eq.~\eqref{eq:Xjdef}. Note that in this regime we have $\vrk<1$ (see Eq.~\eqref{eq:adappmin}). In related works (Refs.~\cite{Agrawal:2017eqm,Machado:2018nqk}) this regime is usually called parametric resonance regime. However, parametric resonance is a generic name for the whole process of particle production due to an oscillating background including the tachyonic regime investigated before. Instead, we call this regime \emph{semi-tachyonic instability} since particle production still occurs in the time interval where $\omega^2<0$. Interestingly the amount of particle production in each oscillation in the semi-tachyonic regime is sensitive to the phase $\theta_{\rm tot}$ in Eq.~\eqref{eq:napp}. One can see that for 
\eq{
\sin\theta_{\rm tot}>\frac{e^{\pi\vrk^2/2}}{\sqrt{1+e^{\pi\vrk^2}}}\,,
}
the number of particles decreases as a result of interference of the solutions before and after scattering (see Ref.~\cite{Kofman:1997yn} for more details). This is the counterpart of the so-called stochastic resonance during preheating in an expanding universe mentioned in Ref.~\cite{Kofman:1997yn}. Although, as mentioned earlier, in our case $\omega ^2$ becomes negative around the point of particle production. Due to the complicated structure of $\theta_{\rm tot}$ it is not straightforward to find an analytic expression for  the total amount of particle production in this regime.

Fig.~\ref{fig:njall} shows the numerical solution of the number density for a longer amount of time until $x_{\rm p}$ and beyond. The red triangles and pink diamonds are the result of analytic computation by the method of successive scattering in Eqs.~\eqref{eq:nj} and \eqref{eq:napp} for both tachyonic and semi-tachyonic regimes respectively. Numerical solution confirms our discussion around Eq.~\eqref{eq:adappmin} about the termination of the tachyonic regime at $x_{\rm et}$ (the first vertical boundary between colored regions). From $x_{\rm et}$ the semi-tachyonic regime starts until $x_{\rm en}$ (the second vertical boundary between colored regions) when we leave the first instability band as discussed around Eq.~\eqref{eq:aen}. Note that while particle production in the tachyonic regime is very efficient and the number of particles grows exponentially within a short time interval, in the semi-tachyonic regime even more amount of particles are produced although within a much longer time. After $x_{\rm en}$ there would be no instability regime and hence no efficient, nonperturbative particle production. After that we might have the conventional perturbative decay. This possibility is discussed in appendix \ref{app:perturbative-decay}.  

\subsection{Backreaction and backscattering}\label{subsec-backreaction}

Thus far we have neglected the effects of the produced dark photons on the dynamics of the axion field $\phi$. As briefly discussed in the introduction and Fig.~\ref{fig:diagram}, the system includes a complicated interplay between dark photons, axion background and its fluctuations. The corresponding equations are solved numerically in the literature \cite{Agrawal:2017eqm,Machado:2018nqk}. However, initially the energy density of dark photons and axion fluctuations are negligible and the dynamics can be studied linearly, corresponding to the production of dark photons (or similarly axion fluctuations) in a homogeneous background of oscillating axion field. This is the implicit assumption of the previous sections. However, this assumption breaks down after significant dark photons are produced by tachyonic instability and one cannot completely trust the analyses of previous sections afterwards.\footnote{In this work we have ignored self-interactions in the axion potential. As a result, the fluctuations of the axion field are not amplified by parametric resonance.} The produced dark photons can affect the evolution of the axion field in at least two ways. The first is the effect of the produced dark photons on the evolution of the background axion field $\bar{\phi}(t)$ that causes its amplitude to decrease faster than $a^{-3/2}$ and as a result resonance is terminated earlier. This effect is usually called the backreaction. The second deals with the inhomogeneities that are induced by the dark photons through which axion acquires momenta even if the initial axion field was homogeneous. This effect is sometimes called the backscattering. In the following we try to estimate when these effects become important from the linear analysis point of view. 

The equation of motion for the background axion field Eq.~\eqref{eq:phibar} can be written approximately as
\eq{
\label{eq:phibar2}
	\ddot{\bar{\phi}}+3H\dot{\bar{\phi}}+m^2\bar{\phi}=\frac{{\alpha}_{a}}{2\f}\ev{E_iB_i+B_iE_i}\,. 
} 
Here we have substituted the quantum expectation value on the right hand side in the Hartree approximation and also used the Weyl ordering in order to get a Hermitian result. By using the definition of the electric and magnetic fields for the dark photons in terms of the mode functions defined in Eq. (\ref{eq:Xcaop}), we obtain 
\eq{
\label{eq:edotb}
\spl{
\frac{1}{2}\ev{E_iB_i+B_iE_i}&=-\frac{1}{4\pi^2a^4}\sum_{\lambda
=\pm}\lambda\int\dd{k}k^3\left[\left(\dot{\chi}-\frac{1}{2}H\chi\right)\chi^*+{\rm cc.}\right]\\
&=\frac{1}{2\pi^2a^4}\sum_{\lambda=\pm}\lambda\int\dd{k}k^3n_{k,\lambda}\left(\sin\psi+\frac{H}{\omega_{\rm ph}}\cos^2(\psi/2)\right)\, .
}
}
In the second line above we have used the solution of Eq.~\eqref{eq:wkbexact} with $\omega^2>0$ to write the mode functions in terms of the number density of the dark photons where we also  defined $\psi \equiv 2\int\dd{t'}\omega_{\rm ph}+\arg\beta-\arg\alpha$. 

The analysis of previous section is valid as long as the right hand side is negligible compared to the terms on the left hand side of Eq.~\eqref{eq:phibar2}. As a result, we can roughly equate them to find an estimate of the time when backreaction becomes important. In order to do this we approximate the integral in Eq.~\eqref{eq:edotb} by evaluating the integrand at $k\approx k_*$ where $k_*=m\kappa_*$ is the peak of the momentum of the spectrum of dark photons given by Eq.~\eqref{eq:k*}. Further, we neglect the subdominant polarization and only consider $\lambda=-1$. Thus, we approximately obtain
\eq{
\label{eq:eb}
\frac{1}{2}\ev{E_iB_i+B_iE_i}\simeq-\frac{k_*^4n_*}{2\pi^2a^4}\,,
}
where $n_{*}$ is the number density of dark photons at $\kappa_*$ and in addition we have neglected the effect of the oscillating factor in Eq. \eqref{eq:edotb}. We denote the time when backreaction effects are important by $x_{\rm br}$ which corresponds to the time when $\frac{{\alpha}_{a}}{2\f}\ev{E_iB_i+B_iE_i}_{\rm br}\,\sim m^2\phi_{\rm os}a_{\rm br}^{-3/2}$. By using the approximate expression Eq.~\eqref{eq:eb} we obtain
\begin{eBox}
\eq{
\label{eq:bcr}
n_*({x_{\rm br}})\sim 2\pi^2 
\Big(\frac{f_{a}}{\alpha_a m}\Big)^2\left(\frac{a_{\rm br}^{5/2}}{\kappa_*\vp}\right)
\sim 2\pi^2 \Big(\frac{f_{a}}{\alpha_a m}\Big)^2\,,
}
\end{eBox} 
where in the second approximate equality we have set $a_{\rm br}\sim\vp^{2/3}$ and the fact that $\kappa_*\sim\vp^{2/3}$.\footnote{In fact since the factor $(f_{a}/\alpha_a m)^2$ is very large, the final result is insensitive to the value of $a_{\rm br}$ used on the right hand side of Eq.~\eqref{eq:bcr}.} By using the above criteria we have numerically solved for the time when backreaction becomes significant. The result is shown in Fig.~\ref{fig:backreaction} by solid curves. As expected, for small values of $\vp$ (or correspondingly small couplings $\alpha_a$) the backreaction effects are not important because particle production is not very efficient. In these cases, we expect that the expansion of the universe plays the leading role in terminating particle production. Further, as the initial energy of the axion background is higher we expect that the backreaction becomes important later. This is also consistent with Fig.~\ref{fig:backreaction} where increasing the value of $\phi_{\rm os}/m$ results in a larger values for $a_{\rm br}$. 

\fg{
	\includegraphics[width=0.63\textwidth]{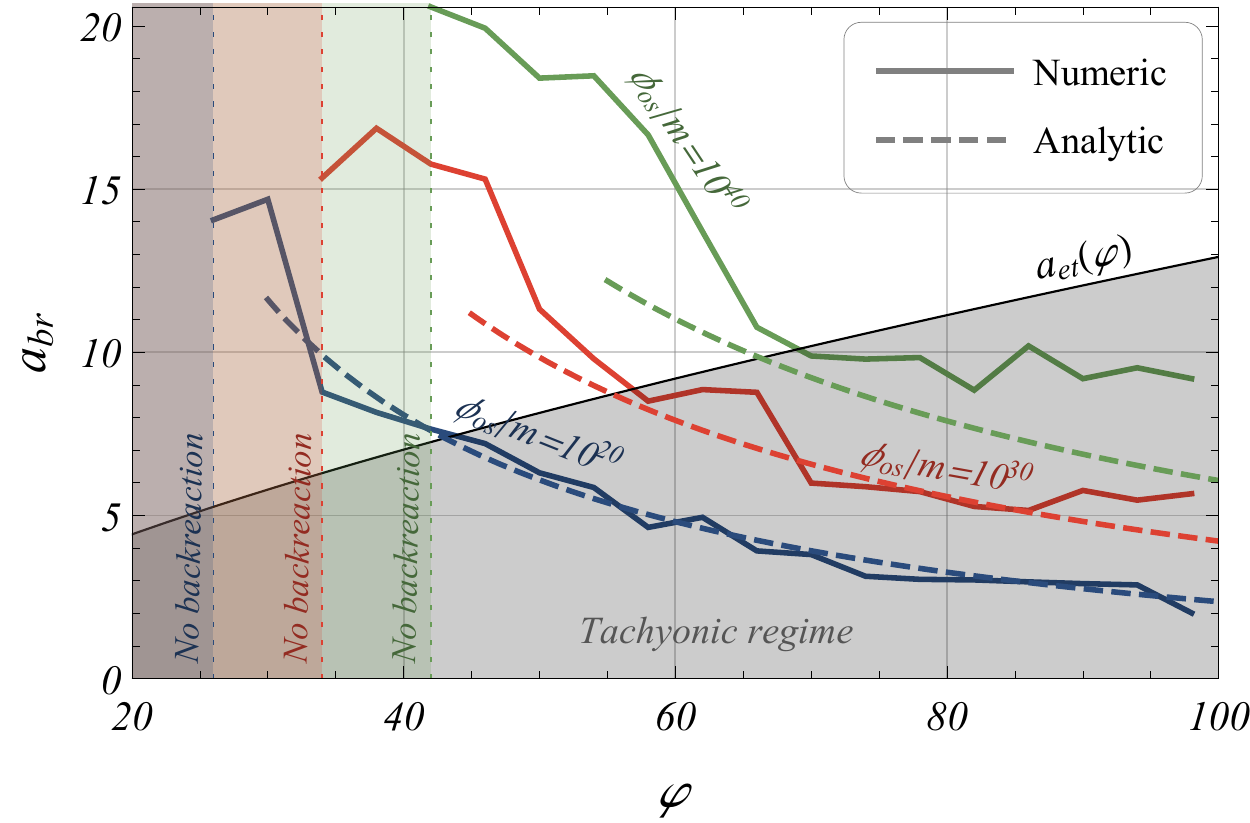}
	\caption{\small Estimation of the time ($a_{\rm br}$) when backreaction effects become important for different choices of parameters $\vp$ and ratio $\phi_{\rm os}/m=\theta f_a/m$. The estimation is based on Eq.~\eqref{eq:bcr}. The solid curves are obtained by solving the dark photons equations numerically while the  dashed curves are the approximate result of Eq.~\eqref{eq:abrapp}. Consistent with one's expectations, for small values of $\vp$ particle production is not very efficient and backreaction effects are not very important. The black curve shows the end of tachyonic regime.}
	\label{fig:backreaction}
}

From the approximate analytic expression for the number of dark photons that we have obtained in the previous section in Eq.~\eqref{eq:sumXapp}, we can obtain an approximate expression for $a_{\rm br}$ as follows
\eq{
\label{eq:abrapp}
a_{\rm br}\approx\frac{\log(\sqrt{2}\pi\frac{\phi_{\rm os}}{m\vp})-\big(0.5\vp^{5/6}-0.95\vp^{1/2}\big)}{\big(0.39\vp^{5/6}-0.57\vp^{1/2}\big)}\,.
} 
This is shown by the dashed curves in Fig.~\ref{fig:backreaction} and seems to be fairly consistent with the numeric solution. Note that from the analysis of the previous section this should only be valid in the tachyonic regime.

Finally, we should also take into account the backscattering effects in which the produced dark photons create axion particles, i.e. generate inhomogeneities in the axion field. By expanding the axion (quantum) field into its Fourier mode we have
\eq{
\label{eq:phik}
\ddot{\phi}_k+3H\dot{\phi}_k+\left(\frac{k^2}{a^2}+m^2\right){\phi}_k=\frac{{\alpha}_{a}}{\f}\big(E_iB_i\big)_k\,. 
} 
As a result, one can write a formal solution by using the Green's function method
\eq{
\phi_k(t)=\frac{{\alpha}_{a}}{\f}\int^t\dd{t'}G_\phi(t,t')\big(E_iB_i\big)_k\,,
}
where $G_\phi$ is the Green's function of the left hand side of Eq.~\eqref{eq:phik}. Ignoring the effects of the axion fluctuations on the dynamics of dark photons, one can use the results of the previous section to compute the power spectrum of these fluctuations, $\ev{\phi_k\phi_{k'}}=(2\pi)^3\Delta_\phi(k)\delta^3(\vb{k}-\vb{k'})$. The backscattering effects can be neglected as long as the energy in the fluctuations of axion is smaller than the energy in its homogeneous part. However, as dark photons are produced this assumption can break down. We denote the time when the backscattering effects become important by $x_{\rm bs}$. As a rough estimation of this time we use the criteria
\eq{
\int\dd[3]{k}\Delta_\phi(x_{\rm bs})\sim\left(\frac{\phi_{\rm os}}{a_{\rm bs}^{3/2}}\right)^2\,.
}  
The computation procedure for  $\Delta_\phi$ is very similar to the one that we will present in the following section for the spectrum of the GWs.\footnote{Main distinctions are different Green's function and polarization factors.} The final result is approximately the same order as  the time when the backreaction becomes important which is given in Eq.~\eqref{eq:bcr}.

It should be noted that the analytic results of the previous sections cannot be employed after the time when the backreaction/backscattering effects become significant and the system must be studied  numerically \cite{Adshead:2015pva,Adshead:2016iae,Peloso:2016gqs,Agrawal:2017eqm,Kitajima:2017peg}. As we have seen above, for some parameters one can neglect these effects and follow the linear analysis of the previous section. However, for the large values of the coupling it is the nonlinear effects that become important. Further, the production of dark photons probably continues for sometime after backreaction/backscattering effects become important. Since our analysis is blind to these effects, in what follows (i) we will assume that nonlinear effects (and not the expansion) terminate particle production and (ii) we use Eq.~\eqref{eq:bcr} to estimate the number density of dark photons. As we will see in the following, we obtain expressions for $\Delta{N}_{\rm eff}$ and $\Omega^{\lambda}_{\rm GW}$ which are consistent with the numerical simulations. 

\subsection{Contribution of dark photons to $N_{\rm eff}$}\label{N-eff}
Although dark photons are not visible, they contribute to the number of relativistic degrees of freedom. Their corrections to the effective number of relativistic degrees of freedom $N_{\rm eff}$ before matter-radiation equality and around the recombination time, will be
\begin{equation}\label{N-eff-NP-def}
\Delta{N}_{\rm eff}
\approx \frac{8}{7}\Big(\frac{11}{4}\Big)^{\frac{4}{3}} \frac{\rho_A}{\rho_\gamma} \,,
\end{equation}
where $\rho_{A}$ is the energy density of dark photons produced non-perturbatively that we discussed in subsection \ref{vector-field-mode-function} and $\rho_{\gamma}$ is the energy density of the standard radiation which is the dominant energy content of the universe in our scenario. We can find an estimate of the correction (\ref{N-eff-NP-def}) in terms of the parameters of the model with our analytic results from previous subsections. The energy density of the produced dark photons can be written in terms of its number density as
\begin{equation}\label{rho-darkphoton-def}
\rho_{A} = \frac{1}{2\pi^2a^4}\sum_{\lambda=\pm} \int dk 
k^3 \, n_{k,\lambda} \,.
\end{equation}
Similar to the previous subsection, we approximate the integral by its value at the peak momentum $k_*$ and as before assume that the backreaction of dark photons terminate particle production. Thus we can approximately write
\begin{equation}\label{rho-darkphoton}
\rho_{A}(a_{\rm br}) \sim \frac{m^4 n_*}{2\pi^2} \sim \Big(\frac{m f_{a}}{\alpha_a}\Big)^2 \,,
\end{equation}
where in the second semi-equality we have substituted the value of $n_*$ from Eq. (\ref{eq:bcr}). Substituting Eq. (\ref{rho-darkphoton}) in \eqref{N-eff-NP-def}, the correction to the effective number of relativistic degrees of freedom at the time of backreaction $x_{\rm br}$ is obtained  to be
\begin{eBox}
\begin{equation}\label{N-eff-NP}
\Delta{N}_{\rm eff} 
\sim
\vp^{2/3}\Big(\frac{\phi_{\rm os}}{{M_{\rm Pl}}}\Big)^2 \,,
\end{equation}
\end{eBox}
where we have substituted $\rho_{\gamma}=3M_{\rm Pl}^2H^2$,  $\Mpl=1/\sqrt{8\pi{G}}\approxeq 2.4 \times 10^{18}$ GeV is the reduced Planck mass, and also we have set $H_{\rm br}\sim m\,\vp^{-4/3}$.

\fg{
	\includegraphics[width=0.53\textwidth]{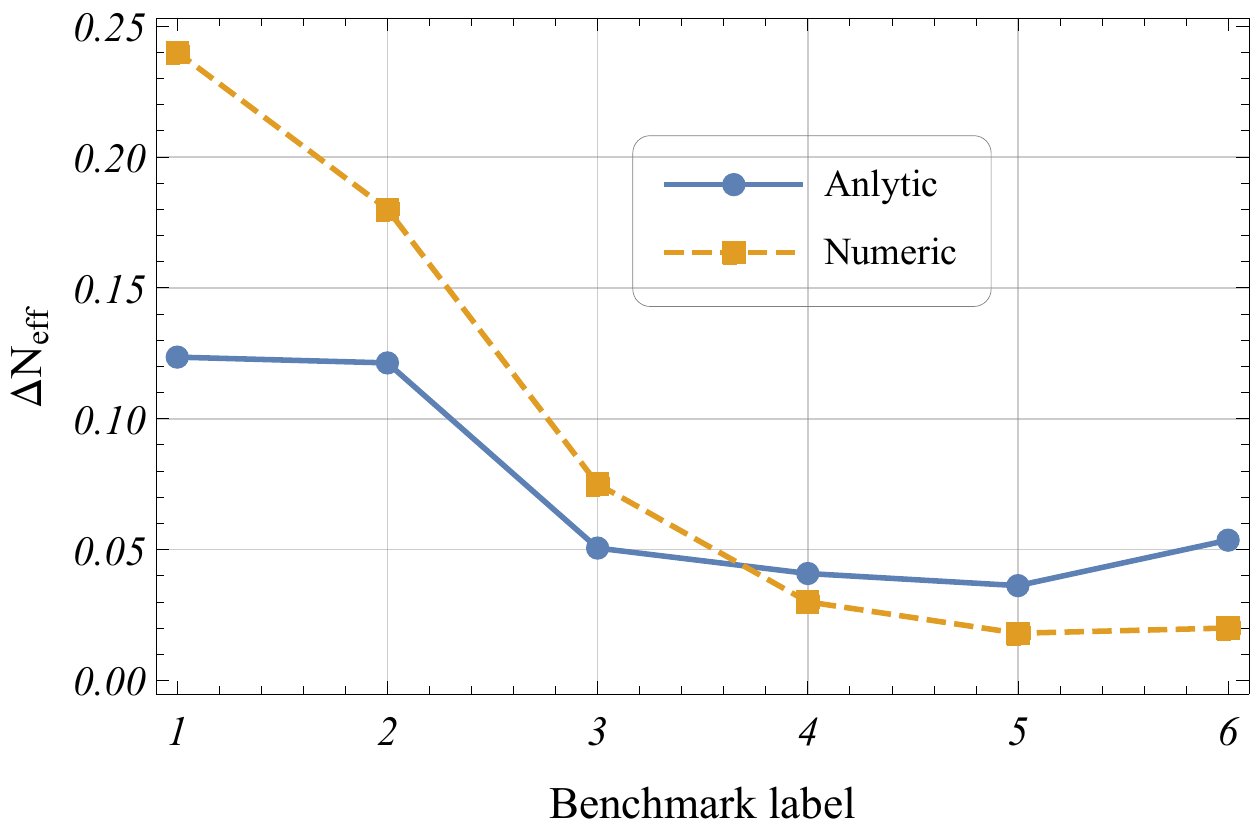}
	\hspace{10mm}
	\raisebox{0.55\height}{\includegraphics[width=0.2\textwidth]{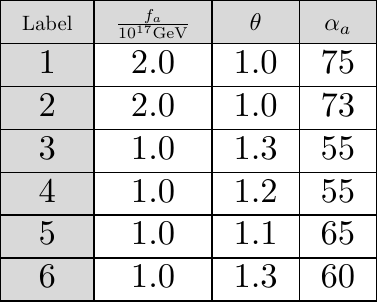}}
	\caption{\small Change in the effective number of relativistic degrees of freedom ($\Delta{N}_{\rm eff}$) due to production of dark photons. The analytic curve is obtained from Eq.~\eqref{N-eff-NP}. The numerical data for different benchmarks is taken from Ref.~\cite{Machado:2018nqk} and their details are presented in the table.}
	\label{fig:dN}
}

We have compared the approximate expression of Eq.~\eqref{N-eff-NP} with the numerical results of Ref.~\cite{Machado:2018nqk} in Fig.~\ref{fig:dN}. They have done the numerical simulation taking into account the backreaction effects but neglected any other effects including the fluctuations in the axion field. The parameters they have used for simulations is presented in the table of Fig.~\ref{fig:dN}. This figure shows that Eq.~\eqref{N-eff-NP} is in qualitative agreement with the numerical studies. The discrepancy is due to the many approximations we have made during the computation such as neglecting the subdominant dark photon polarization, neglecting dark photons with momenta far from $k_*$ and neglecting dark photons produced after the time of backreaction. Improvement of those approximations should result in even better agreement with the numerical results. 

Before concluding this section we note that the current bound on the change in the effective number of relativistic degrees of freedom is $\Delta{N}_{\rm eff}<0.3$ \cite{Aghanim:2018eyx}. Fig.~\ref{fig:dN} shows that the model under consideration can be consistent with observations.

\section{Production of gravitational waves}\label{sec-GWs}

In the previous section, we have seen that in the presence of Chern-Simons coupling, considerable amount of energy can be extracted from the background axion field by tachyonic instability, and dark photons then will be  produced mostly with a specific energy (or equivalently momentum) denoted by $k_*=m\kappa_*$ where $\kappa_*$ is given by Eq.~\eqref{eq:k*}. As the dark sector is minimally coupled to the gravity sector, the produced dark photons provide a nonlinear source for the linear tensor perturbations in the metric sector which are defined as $\delta{g}_{ij} = a^2 h_{ij}$ subject to the traceless and transverse conditions $h_i{}_i=0=\partial_ih_{ij}$ (see \cite{Caprini:2018mtu} for a review of GW in cosmology). 

One may also consider scalar perturbations in both dark and gravity sectors. Note that we have assumed a homogeneous configuration for the axion field. However, as we have discussed in subsection \ref{subsec-backreaction}, inhomogeneities in the axion field will show up due to the backscattering of the produced dark photons even if we start with a completely homogeneous axion field at the beginning. Indeed these inhomogeneities in the axion field will not affect curvature perturbations during the RD era as the axion background energy density is negligible  compared to the standard radiation energy density. In this respect, the inhomogeneities in the axion field can be thought of as isocurvature modes. Of more importance is however the nonlinear effects of these scalar modes on the linear evolution of the tensor modes which was the subject of some very recent studies \cite{Cai:2018dig,Hajkarim:2019nbx,Domenech:2020kqm,Pi:2020otn}. Moreover, as was mentioned in subsection \ref{subsec-backreaction}, we only focus on the case in which the backreaction terminates the process of particle production in a very short interval in the tachyonic resonance regime. Since from the lattice simulations we know that nonlinear backscattering effects are negligible in this regime, we will not take into account the scalar perturbations in the axion field. Indeed, the main features of the produced GWs will survive through our linear analysis. In this regards, in this section we only look for the effects of the dark photons as a source for GWs.

The tensor modes $h_{ij}$ can be expressed as a combination of right or left handed polarization as follows 
\eq{
h_{ij}(\xv,t)=\sum_{\lambda=\pm}\int\frac{\dd[3]{k}}{(2\pi)^{3/2}}e_{ij}^\lambda(\hat{\kv})h_\lambda(k)e^{i\kv.\xv}\,,
}
where $e_{ij}^\lambda(\hat{\kv})$ are circular polarization tensors properties of which are presented in appendix \ref{app:pol}. Considering the dark sector Lagrangian density (\ref{L-axion}) minimally coupled to the Einstein-Hilbert Lagrangian density, it is straightforward to show that the equations of motion for $h_\lambda(k)$ is
\eq{
\label{eq:tensor}
\ddot{h}_\lambda+3H\dot{h}_\lambda+\frac{k^2}{a^2}h_\lambda=\frac{2}{\Mpl^2}\Pi_\lambda\,,
}
where $\Pi_\lambda \equiv e_{ij}^{-\lambda}T^{ij}(\kv)$ and $T_{ij}(\kv)$ is the Fourier transform of the spatial part of the electromagnetic energy-momentum tensor given by
\eq{
\label{eq:emtens}
T_{ij}=-a^2(E_iE_j+B_iB_j)+ \frac{1}{2} a^2 (E_\ell E_\ell+B_\ell B_\ell) \delta_{ij} \,.
}
The physically interesting object is the GW energy density per logarithmic wave number, $d\rho_{\rm GW}/d\log k$, related to the energy density $\rho_{\rm GW}$ of GWs in the form
\eq{
\rho_{\rm GW}=\frac{\Mpl^2}{4}\langle\dot{h}_{ij}\dot{h}_{ij}\rangle\equiv\int\frac{\dd{k}}{k}\dv{\rho_{\rm GW}}{\log k}\,,
}
where $\langle \,  \rangle$ means suitable temporal and spatial averaging \cite{Caprini:2018mtu}. To simplify the derivations, in what follows 
we use conformal time $\eta$ but  in order to use the results of the previous section  
we rewrite all results at the end in terms of the cosmic time. In order to obtain the GW energy density, we look at the power spectrum $\Delta(k,\lambda,\eta)$ of the GWs defined by the two-point correlation function of the tensor polarizations 
as follows 
\eq{
\langle h_\lambda(\kv,\eta)h^*_{\lambda'}(\kpv,\eta)\rangle=\frac{2\pi^2}{k^3}\delta_{\lambda\lambda'}\delta^3(\kv-\kpv)\Delta(k,\lambda,\eta)\,,
}
where we have used the translation and rotation symmetry of the FLRW universe to ensure the proportionality to delta functions in the absence of any net polarization \cite{Caprini:2018mtu,Weinberg:2008zzc}. Note that the parity violating interaction shows itself through the explicit polarization dependence of $\Delta$. The solution for the source free tensor modes, i.e. Eq.~\eqref{eq:tensor} in the RD universe with $\Pi_\lambda=0$, is easily obtained to be
\eq{
\label{eq:hfree}
h_\lambda=\frac{1}{a(\eta)} \big( A_\lambda e^{ik\eta}+B_\lambda e^{-ik\eta} \big)\,,
}
where $A_\lambda$ and $B_\lambda$ are two constants. For this solution, it is easy to show that in the sub-horizon limit $k\gg aH$, 
\eq{
\dv{\rho^\lambda_{\rm GW}}{\log k}\simeq\frac{\Mpl^2k^2}{4a^2}\Delta(k,\lambda,\eta)\,,
}
in which $\Delta\propto\ev{|A_\lambda|^2+|B_\lambda|^2}$ where we have ignored oscillatory terms which vanish after taking the temporal average. For our case with the source, we can write the solution by means of the Green's theorem as
\eq{
h_\lambda(k,\eta)=\int_{\eta_i}^{\eta}\dd{\eta'}G_h(\eta,\eta')\Pi_\lambda(k,\eta')\,,
}
where $\eta_i$ is the initial conformal time at which the source is turned on, i.e. corresponding to the time when the axion field starts to oscillate. The retarded Green's function of Eq.~\eqref{eq:tensor} in terms of the conformal time is given by $G_h(\eta,\eta')=(a(\eta')/a(\eta))\sin(k(\eta-\eta'))/k$. We assume that before $\eta_i$ the solution to the homogeneous equation vanishes\footnote{This is not quite right since for example we have tensor modes from inflation or other sources. We assume that they are subdominant and that cross correlations among different sources vanish after suitable temporal/spatial averaging.}. Since the vector modes start to decay soon after the parametric 
amplification, we can safely assume that the source $\Pi_\lambda$ is nonzero only until the time $\eta_f$. After that ($\eta\gg\eta_f$), the solution can be expressed in the form of Eq.~\eqref{eq:hfree} and the coefficients $A_\lambda$ and $B_\lambda$ can be computed by matching the solutions. As a result, the amplitude of the two-point function of GWs at late time then takes the following form
\eq{
\Delta(k,\lambda,\eta)=\frac{k}{\Mpl^4\pi^2a(\eta)^2}\int_{\eta_i}^{\eta_f}\dd{\eta_1}\dd{\eta_2}a_1a_2\cos(k(\eta_1-\eta_2))\langle\Pi_\lambda(k,\eta_1)\Pi^*_\lambda(k,\eta_2)\rangle'\,,
}
where a prime over the unequal-time average means that the delta functions are not included.  To compute the unequal-time average, we need to express $\Pi_\lambda$ in terms of the mode functions which were obtained in the last section. By using Eq.~\eqref{eq:emtens},  and from the definition of electric and magnetic fields for the dark photons, after some algebra we get
\eq{
\Pi_\lambda(k,\eta)=-\frac{1}{a}\sum_{r,r'=\pm}\int\frac{\dd[3]{p}}{(2\pi)^{3/2}}\mathcal{P}^{r,r'}_\lambda\mathcal{M}_{r,r'}(p,p',\eta) \,,
}
where we have defined $\vb{p}' \equiv \kv-\vb{p}$. The first part of the integrand above, $\mathcal{P}^{r,r'}_\lambda$,  comes from the polarization vectors and tensors as
\eq{
\label{eq:P}
\mathcal{P}^{r,r'}_\lambda=\frac{1}{4}(1+r\lambda\cos\gamma)(1+r'\lambda\cos\gamma')\,,
}
where $\cos\gamma \equiv \hat{\kv}.\hat{\vb{p}}$ and $\cos\gamma' \equiv \hat{\kv}.\hat{\vb{p}}'$. 
The second part of the integrand above, $\mathcal{M}_{r,r'}$,  is constructed from the mode functions as 
\eq{
\mathcal{M}_{r,r'}=\bigg[\dot{X}_r(p)-HX_r(p)/2\bigg]\bigg[\dot{X}_{r'}(p')-HX_{r'}(p')/2\bigg]+rr'\frac{pp'}{a^2}X_r(p)X_{r'}(p')\,,
}
where $X_r=\chi_{r}a_{\kv,r}+\chi_{r}^*a^\dagger_{-\kv,r}$ and $\chi_r(t)$ are the mode functions for the dark photons. The derivation of the mode functions was one of the main focuses of the previous section. Note that the explicit dependence on the GW polarization $\lambda$ comes only through the term $\mathcal{P}^{r,r'}_\lambda$. 

By using the Wick's theorem and keeping only the connected part, we obtain
\eq{
\langle\Pi_\lambda(k,\eta)\Pi^*_\lambda(k,\eta')\rangle'=\frac{2}{aa'}\sum_{r,r'}\int\frac{\dd[3]{p}}{(2\pi)^{3/2}}(\mathcal{P}^{r,r'}_\lambda)^2\left(\upsilon\upsilon'+rr'\frac{pp'}{a^2}\chi\chi'\right)_{\eta}\left(\upsilon\upsilon'+rr'\frac{pp'}{a^2}\chi\chi'\right)_{\eta'}^*\,,
}
where we have defined $\upsilon_{p,r} \equiv \dot{\chi}_{p,r}-H\chi_{p,r}/2$ (note that the dot is the derivative with respect to the cosmic time) and the prime over the mode functions means that the corresponding momentum and polarization are $p'$ and $r'$ respectively. The subscripts $\eta$ and $\eta'$ of the parentheses denote that the expressions are evaluated at the corresponding conformal times. 

Using the above relations and after some re-arrangements, we find
\eq{
\label{eq:drhoGW}
\Omega^\lambda_{\rm GW}=\frac{k^3}{6\pi^2\Mpl^4a^4H^2}\sum_{r,r'}\int\frac{\dd[3]{p}}{(2\pi)^{3/2}}(\mathcal{P}^{r,r'}_\lambda)^2\left[|I_c|^2+|I_s|^2\right]\,,\qquad \Omega^\lambda_{\rm GW} \equiv \frac{1}{\rho_{\rm c}} \dv{\rho^\lambda_{\rm GW}}{\log k} \,, 
}
where we have defined the dimensionless fractional GW energy density $\Omega^\lambda_{\rm GW}$ and $\rho_{\rm c}=3\Mpl^2H^2$ is the critical energy density of the universe. In the above expression for the fractional GW energy density we have defined
\eq{
\label{eq:Ics}
I_{\left(\substack{c \\ s}\right)} \equiv \int_{\eta_i}^{\eta_f}\dd{\eta}
\begin{pmatrix}
\cos k\eta\\
\sin k\eta
\end{pmatrix}\left[\upsilon_{p,r}\upsilon_{p',r'}+rr'\frac{pp'}{a^2}\chi_{p,r}\chi_{p',r'}\right]\,.
}
We are interested in obtaining an analytic expression for Eq.~\eqref{eq:drhoGW} at the present time. In this regard, we have to impose  a couple of approximations in various steps. The first approximation we make is to replace the spectrum of dark photons by an exact delta function in momentum space located at $k_*$ ($=m\kappa_*$) and also to take into account only the left-handed polarization. This is a reasonable approximation according to the discussion of the last section that the spectrum of photons is sharply peaked around $k_*$ and that one of the two polarizations is dominant. This approximation helps us to get rid of the momentum integration and the sum over polarizations in Eq.~\eqref{eq:drhoGW}. More explicitly let us write the integral measure in the momentum space as
\eq{
\dd[3]{p}=\frac{pp'}{k}\dd{p}\dd{p'}\dd{\Phi}\,,
}
where $\Phi$ is the azimuthal angle with trivial integration yielding $2\pi$. Also after trivial algebra, we find
\eq{
\cos\gamma=\frac{k^2-p'^2+p^2}{2kp}\,,\qquad\cos\gamma'=\frac{k^2-p^2+p'^2}{2kp'}\,,
}
which can be used in Eq.~\eqref{eq:P}. Following the above approximation, we can set $p\simeq p'\simeq k_*$ and ignore the contribution from the dark photons with other momenta. In this respect, for a general momentum $k$ of the GWs,  we have
\eq{
\cos\gamma=\cos\gamma'=\frac{k}{2k_*}\,.
}   
In other words, the three vectors $\vb{p}$, $\vb{p}'$ and $\kv$, which must satisfy the momentum conservation, form an isosceles triangle (see Fig.~\ref{fig:mmcons}). Note that based on this approximation we must have $k\lesssim2k_*$ and modes with higher momenta cannot be produced. Further, considering the fact that one of the polarizations is dominant (in our convention it is chosen to be the left-handed one) we set $r=r'=-1$. As a result, Eq.~\eqref{eq:drhoGW} can be  written approximately as
\eq{
\label{eq:dOmega}
\Omega^\lambda_{\rm GW} \simeq\left[\frac{k_*^4\Delta k_*^2}{96\pi^4\Mpl^4a^4H^2}\right]\left[\left(\frac{k}{2k_*}\right)^2\left(1-\lambda\frac{k}{2k_*}\right)^4\right]\bigg[|I^{--}_{c*}|^2+|I^{--}_{s*}|^2\bigg]\,,\quad k\lesssim2k_*\,,
} 
where a star in the subscript of $I_c$ and $I_s$ indicates that we have set $p=p'=k_*$, the superscript ``$--$'' means that we have set $r=r'=-1$, and $\Delta k_*$ is the width of the peak in the spectrum of the produced photons which can be approximated as $\Delta k_*\sim k_*$.

\fg{
\includegraphics[width=0.3\textwidth]{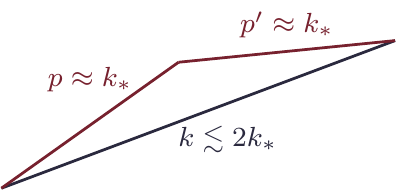}
\caption{\small By momentum conservation we must have $\vb{k}=\vb{p}+\vb{p'}$ where $\vb{k}$ is the momentum of GW and $\vb{p}$ and $\vb{p'}$ are momenta of dark photons. The main contribution of dark photons come from momenta $p=p'=k_*$ which are significantly produced in tachyonic regime. As a result, the momentum of GW can be at most $2k_*$ corresponding to a folded triangle.}
\label{fig:mmcons}
}

The remaining part is to take the time integral appearing in Eq.~\eqref{eq:Ics}. Based on the approximation $p=p'=k_*$ and $r=r'=-1$ that we use here, it takes the form
\eq{
\label{Ics}
I_{\left(\substack{c \\ s}\right)}\simeq\int_{\eta_i}^{\eta_f}\dd{\eta}
\begin{pmatrix}
\cos k\eta\\
\sin k\eta
\end{pmatrix}\bigg[\dot{\chi}^2+\frac{k_*^2}{a^2}\chi^2\bigg]\,,
}
where we have neglected a term with the Hubble expansion rate in $\upsilon_{r,p}\simeq\dot{\chi}_{r,p}$. Plugging the general solution of the mode functions from Eqs.~\eqref{eq:wkbexact} and \eqref{eq:wkbexact2} in Eq.  (\ref{Ics}) we obtain 
\eq{
\label{eq:Ialpha}
I_{\left(\substack{c \\ s}\right)}=\int_{y_i}^{y_f}\frac{\dd{y}}{2\omega_*}
\begin{pmatrix}
\cos \kappa y\\
\sin \kappa y
\end{pmatrix}
\bigg[\left(\frac{\kappa_*^2}{a^2}-\omega_*^2\right)\left(\alpha^2e^{-2i\int\dd{x}\omega_*}+\beta^2e^{2i\int\dd{x}\omega_*}\right)+2\alpha\beta\left(\frac{\kappa_*^2}{a^2}+\omega_*^2\right)\bigg]\,,
}
where we have defined the rescaled conformal time $y\equiv \eta/m$ and the dimensionless GW momentum $\kappa=k/m$. Further, $\kappa_*$ is given by Eq.~\eqref{eq:k*} and by $\omega_*$ we mean that we have set the momentum of the dark photon to $\kappa_*$ in Eq.~\eqref{eq:omega2new}. 

Note that the dependency on the momentum of GWs appeared only through the argument of sine/cosine function. From the definition of the number of particles we have $|\alpha|^2\simeq|\beta|^2\simeq|\alpha\beta|\simeq n_*$ which, from the previous section, we know that it grows exponentially due to tachyonic and semi-tachyonic particle productions. As a result, we expect that the value of the integral is dominated by the late time behaviour of its integrand which is exponentially larger than its early values. That is the reason why in writing Eq.~\eqref{eq:Ialpha} we have only considered the solution which is valid only for $\omega^2>0$. Since due to the backreaction the number of particles does not grow at late time, we can approximately pull it outside of the integral yielding 
\eq{
\label{eq:Ialpha2}
\spl{
I_{\left(\substack{c \\ s}\right)}\simeq n_*\int_{y_i}^{y_f}\frac{\dd{y}}{2\omega_*}
\begin{pmatrix}
\cos \kappa y\\
\sin \kappa y
\end{pmatrix}
\bigg[&\left(\frac{\kappa_*^2}{a^2}-\omega_*^2\right)\left(e^{2i\arg\alpha-2i\int\dd{x}\omega_*}+e^{2i\arg\beta+2i\int\dd{x}\omega_*}\right)\\
&+2e^{i\arg\alpha+i\arg\beta}\left(\frac{\kappa_*^2}{a^2}+\omega_*^2\right)\bigg]\,.
}
}
Note that the amplitude of $I_c$ or $I_s$ is mainly determined by the prefactor $n_*$. The integrand is of oscillating nature and can be approximately computed by the method of stationary phase. In the regime where particle production is no longer active, the phases $\arg\alpha$ and $\arg\beta$ are approximately constant. As a result, the term in the second line of Eq.~\eqref{eq:Ialpha2}  vanishes due to the oscillations of the sine/cosine functions. The nonvanishing terms that result in saddle point(s) are of the form
\eq{
\label{eq:saddleint}
\int\frac{\dd{y}}{2\omega_*}\left(\frac{\kappa_*^2}{a^2}-\omega_*^2\right)e^{\pm i\Psi}\,,\qquad\Psi  \equiv \kappa y-2\int\dd{x}\omega_*\,.
} 
The saddle points are determined as usual by the equation ${\rm d}\Psi/{\rm d}x$ which results in
\eq{\label{saddle-points}
\frac{\sin x}{\sqrt{a(x)}}=\frac{4\kappa_*^2-\kappa^2}{4\vp\kappa_*}\,.
}

\fg{
	\includegraphics[width=0.63\textwidth]{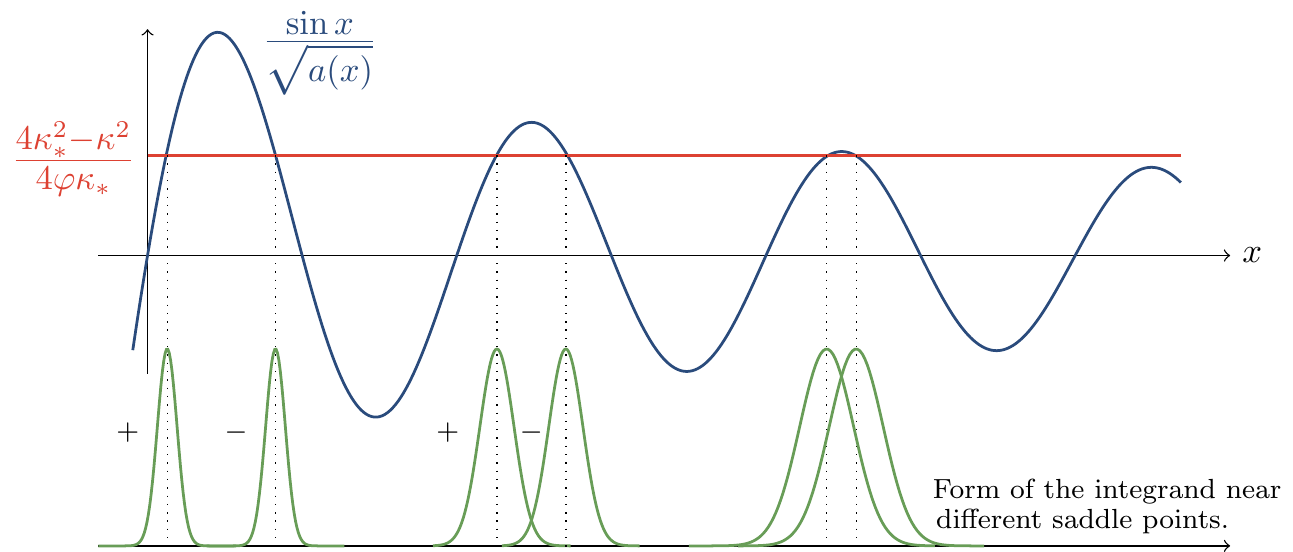}
	\caption{\small Different solutions of the saddle equation \eqref{saddle-points} are shown on the top. In the bottom, the behaviour of the integrand around these saddle points is depicted. The two solution must be accounted as a distinct saddle point as long as they are well separated. Distinct saddles that appear around each maximum of the sine function contribute with different signs thus approximately cancel each other.}
	\label{fig:saddle}
}

Note that from the momentum conservation we have $\kappa<2\kappa_*$ and also $\kappa_*<\vp$, thus the right hand side of the above equation is always between zero and unity which means that the above equation has many solutions as the sine function oscillates. This is shown in Fig.~\ref{fig:saddle}. As a result, to compute the integral we need to sum over all the saddle points. Fortunately, most of these contributions cancel each other as justified below. 

We label the solutions of Eq. \eqref{saddle-points} by $x^\pm_{i}$ where $i$ enumerates the number of oscillations of the sine function while $\pm$ shows that we have two solutions located in either side of the maximum of the sine function (see Fig.~\ref{fig:saddle}). However most of these saddle points do not contribute to the final result. The reason is that for the early solutions the contribution from the point $x_i^-$ cancels the contribution from the successive point $x_i^+$ on the other side of the maximum since the sign of the term ${\rm d}^2\Psi/{\rm d}x^2$ corresponding to $x_i^-$ and $x_i^+$ are opposite to each other so they contribute to the final result with different signs (see for example Ref.~\cite{arfken}). Here, we have neglected the difference in the value of the scale factor corresponding to to $x_i^-$ and $x_i^+$ (similar approximation was exploited around Eq.~\eqref{eq:xjpm}). Thus, contributions from the two successive saddle points $x_i^-$ and $x_i^+$ cancel each other at early times and we do not need to consider them. 

In the course of time, the successive saddle points $x_i^-$ and $x_i^+$ becomes closer and closer since the amplitude decreases due to the expansion and at some point $x_i^-$ and $x_i^+$  become so close to each other (and to the maximum of the sine function).  Besides, the width of the Gaussian factor by which we approximate the contributions from each saddle point is proportional to $1/\sqrt{{\rm d}^2\Psi/{\rm d}x^2}$ and grows with the scale factor. As a result, the two solutions $x_i^-$ and $x_i^+$ cannot be treated as two separate saddle points. Based on these considerations, we can only consider the solution to Eq.~\eqref{saddle-points} for which we have $\sin x\approx1$.
 
This is a unique solution denoted by $x_s$ and with good accuracy gives the dominant contribution to the integral with properties
\eq{
\label{saddledata}
a_s=\left(\frac{4\vp\kappa_*}{4\kappa_*^2-\kappa^2}\right)^2\,,\qquad\dv[2]{\Psi}{x}\Bigg\rvert_s=\frac{1}{a_s^3}\frac{4\kappa_*^2-\kappa^2}{8\kappa}\,.
}     
Then the value of the integral in Eq.~\eqref{eq:saddleint} is computed to be
\eq{
\label{saddledataint}
\int\frac{\dd{y}}{2\omega_*}\left(\frac{\kappa_*^2}{a^2}-\omega_*^2\right)e^{\pm i\Psi}\simeq\sqrt{\frac{\pi}{\kappa}}\frac{(\kappa_*^2-\kappa^2/4)^{3/2}}{\vp\kappa_*}e^{\pm i\Psi_s}\,,
}
where $\Psi_s$ is the value of the phase at the dominant saddle point. 

Plugging the above result into the integral of Eq.~\eqref{eq:Ialpha2} and then using the outcome in Eq.~\eqref{eq:dOmega}, we obtain the following result for the GWs at the time of emission
\eq{
\label{eq:omgw}
\Omega^{\lambda}_{\rm GW,em}\simeq\left[\frac{k_*^6}{192\pi^3\Mpl^4a_{\rm em}^4H_{\rm em}^2}\frac{n_*^2\kappa^3_*}{\vp^2}\right]\left[\left(\frac{k}{2k_*}\right)\left(1-\lambda\frac{k}{2k_*}\right)^4\left(1-\frac{k^2}{4k^2_*}\right)^3\right]\,,
}
where the subscript ${\rm em}$ denotes the value of the corresponding quantity at the time of emission. Since GWs are produced in RD era we can write $a_{\rm em}^4H_{\rm em}^2 \simeq a_{\rm os}^4 H_{\rm os}^2 \simeq m^2/4$ as we have set $a_{\rm os}=1$ and $m=2H_{\rm os}$. The amplitude of the GWs also depends on the number of produced photons $n_*$. As we already asserted we use Eq.~\eqref{eq:bcr} as an estimate of $n_*$ at the time backreaction becomes important. Furthermore, the value of $k_*$ can be computed from Eq.~\eqref{eq:k*} in terms of the parameters of the setup. Putting these all together, we find the following expression for the GW energy density at the time of emission
\begin{eBox}
\eq{
	\label{eq:omgw-emission}
	\Omega^{\lambda}_{\rm GW,em} \simeq
	\left[\frac{2}{243\pi} \left(\frac{\phi_{\rm os}}{M_{\rm Pl}}\right)^4\right]
	\left[\left(\frac{k}{2k_*}\right)\left(1-\lambda\frac{k}{2k_*}\right)^4\left(1-\frac{k^2}{4k^2_*}\right)^3\right]\,.
}
\end{eBox}

\fg{
	\includegraphics[width=0.73\textwidth]{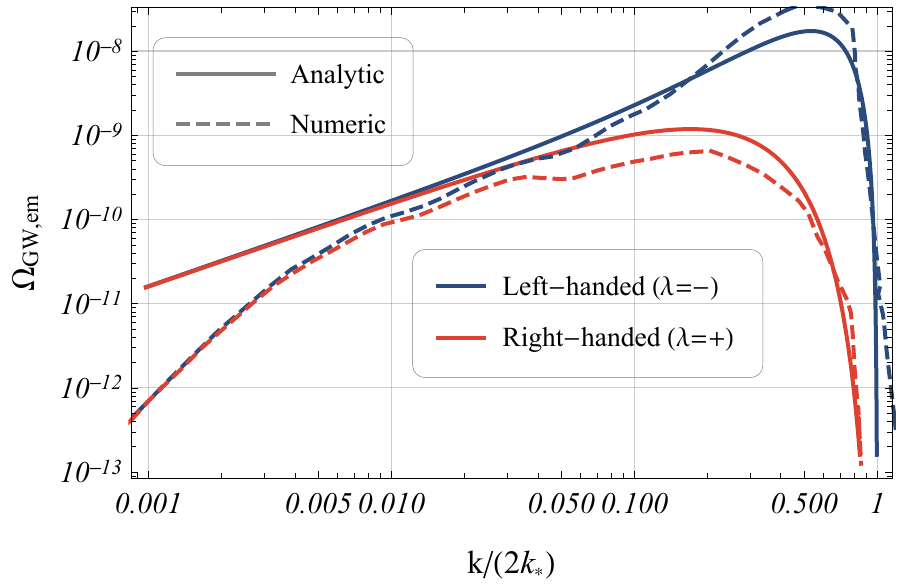}
	\caption{\small Spectrum of the GW for both polarizations produced by dark photons. The solid curves show the analytic spectrum we obtained in Eq.~\eqref{eq:omgw-emission} while the dashed curves show the corresponding numerical results found in Ref.~\cite{Machado:2018nqk}. The numeric data is extracted from Fig.~3 of Ref.~\cite{Machado:2018nqk} where the parameters are $\alpha_{a}=55$, $\theta=1.2$ and $\f=10^{17}$GeV. This corresponds to $\phi_{\rm os}=\theta\f=1.2\times10^{17}$GeV and $\vp=\alpha_{a}\theta=66$. In extracting the numerical data we have re-scaled the horizontal axes such that the peak of the analytic and numeric spectra happens at the same $k/2k_*$. The numerical value needed for a perfect match is $k_*^{\rm num}=1.8\,m\kappa_*$ where $\kappa_*$ is given in Eq.~\eqref{eq:k*}.}
	\label{fig:GW2}
}

The spectrum of GW is depicted in Fig.~\ref{fig:GW2}. The solid blue and red curves are analytic results of Eq.~\eqref{eq:omgw-emission} for left and right handed GWs respectively. The sharp suppression for $k\gtrsim2k_*$ is because we have included only photons with momenta $\sim k_*$. For comparison we have also extracted the numerical data of the spectrum from Ref.~\cite{Machado:2018nqk} for $\alpha_{a}=55$, $\theta=1.2$ and $\f=10^{17}$GeV. For these parameters the analytic results of previous sections predict the value of $k_*=m\kappa_*$ with $\kappa_*\approx12.4$ obtained from Eq.~\eqref{eq:k*}. However, the peak of the numerical data is larger than this value by a factor less than 2 for this case. In fact the perfect match in Fig.~\ref{fig:GW2} is obtained by re-scaling the numerical data with $k_*^{\rm num}=1.8\,m\kappa_*$. We trace this discrepancy back to the numerous approximations we have made during the computation.        

Fig.~\ref{fig:GW2} shows that our analysis provides a good approximation for the amplitude of the GW, the position of the peak (up to an $\order{1}$ factor) as well as its spectrum specially in the vicinity of the peak. However, it should be noted that the analytic expression Eq.~\eqref{eq:omgw-emission} cannot be trusted in the IR limit, i.e. for very long wavelengths. The main reason is that for very small momenta, the saddle point approximation fails as the resulting integral is divergent which can be seen from Eqs.~\eqref{saddledata} and \eqref{saddledataint}. In fact, it is not difficult to see that in this limit the integral appearing in Eq.~\eqref{eq:drhoGW} is actually independent of $k$. As a result, one expects the momentum dependence of the IR tail be of the form $\Omega_{\rm GW}\propto k^3$. This is a universal behaviour for any casual mechanism of GW production \cite{Caprini:2009fx,Hook:2020phx}. This behaviour applies for modes with wavelength $k^{-1} \gg T$ where $T$ is the typical time or length scale of GWs production. In our setup, we have only focused on the growing modes for which the typical time scale for tachyonic growth is $T\propto m^{-1}$ as we explained in the paragraph after Eq. \eqref{eq:k*}. As a result, the IR behaviour can be seen for modes $k \ll m$ (corresponding to $(k/2k_*)\ll\vp^{-2/3}$). This is the region where our analytic expression deviates from the numerical results shown in Fig.~\ref{fig:GW2}.

Note that in expression Eq.~\eqref{eq:omgw-emission}, the dependence on the coupling ${\alpha}_{a}$ is very weak and only comes in through the value of $k_*$. Instead,  the main dependence is on the initial value of the axion field which is usually written in the form $\phi_{\rm os}=\theta\f$. While Fig.~\ref{fig:GW2} suggests that what we have here is fairly consistent with the numerical result, it is parametrically different from what has been suggested by Refs.~\cite{Machado:2018nqk,Machado:2019xuc}. It seems that both derivations have their own pitfalls and do not have enough numerical data to discriminate between them.   
	
The energy density of GW at present time can be computed by taking into account the effects of decoupling of the relativistic particles during the course of expansion, and we have \cite{Caprini:2018mtu}
\eq{
	\label{eq:omgw-today}
	\Omega^{\lambda}_{\rm GW,0} = \Omega^{\lambda}_{\rm GW,em}
    \left(\frac{g_{{\rm s,}0}}{g_{{\rm s,em}}}\right)^{4/3} \left(\frac{T_0}{T_{\rm em}}\right)^{4} \left(\frac{H_{\rm em}}{H_0} \right)^2 
    = \Omega^{\lambda}_{\rm GW,em}
    \left(\frac{g_{{\rm s,}0}}{g_{{\rm s,em}}}\right)^{4/3} \left(\frac{g_{\rho,{\rm em}}}{g_{\rho,0}}\right) 
    \Omega_{\gamma,0} 
    \,,
}
where $g_{\rm s}$ and $g_\rho$  are the effective numbers of relativistic degrees of freedom associated to the entropy and the energy density respectively, the subscripts $0$ and ${\rm em}$ show the corresponding values at the present time and at the time of emission respectively, and $ \Omega_{\gamma,0}$ is the present value of the fractional dimensionless energy density of radiation. Considering the typical values of $g_{{\rm s,}0} = 3.91$ \cite{Kolb:1990vq}, $g_{{\rm s,em}}\simeq g_{\rho,{\rm em}} =106.8$, $g_{\rho,0}=2$, $\Omega_{\gamma,0} = 5.38\times10^{-5}$ \cite{Tanabashi:2018oca}, the GW energy density today can be obtained from \eqref{eq:omgw-emission} and \eqref{eq:omgw-today} as 
\eq{
	\label{eq:omgw-today-f}
	\Omega^{\lambda}_{\rm GW,0} \simeq {\cal O}\big(10^{-7}\big) \left(\frac{\phi_{\rm os}}{\Mpl}\right)^4
	\left[\left(\frac{k}{2k_*}\right)\left(1-\lambda\frac{k}{2k_*}\right)^4\left(1-\frac{k^2}{4k^2_*}\right)^3\right] \,.
}
It should be noted that the analytic results for the GW spectrum in Eq.~\eqref{eq:omgw-emission} or (\ref{eq:omgw-today-f}) can be used as templates in GW data analysis to extract signals. Further, one can easily relate the observed spectrum to the fundamental parameters of the theory by fitting the values of $k_*$ and $\phi_{\rm os}$. 

Finally, Eq.~\eqref{eq:omgw-emission} shows that the chirality of the dark photons has resulted in a polarized GW spectrum which is the consequence of the parity violating nature of the Chern-Simons interaction. Interestingly, astrophysical stochastic GWs are expected to be unpolarized. Hence it is in principle possible to discriminate the polarized GW signal we have obtained in this model from the unpolarized astrophysical sources by the GW detectors. In this regard, the parity violation can be measured following the method suggested in Refs. \cite{Seto:2006hf,Seto:2006dz,Seto:2007tn,Seto:2008sr,Smith:2016jqs,Domcke:2019zls,Crowder:2012ik}.

\section{Summary and conclusions}\label{sec-conclusion}

For the range of masses $m\gtrsim10^{-27}$eV, axion-like fields  are expected to  oscillate in the radiation dominated era when the Hubble expansion rate drops below their mass scale. The natural interaction between the axion-like field and the gauge bosons is through the Chern-Simons coupling. If the axion-like field is coupled to the dark gauge bosons, the coupling constant can be large enough that sizeable amounts of energy can be 
transferred from the axion-like field to the dark gauge boson through the tachyonic resonance process. These types of scenarios are well studied in both the early universe setups like inflation and late time scenarios like axionic vector dark matter scenarios. 

Some interesting results arise when we consider interaction of the axion-like field with the hidden sector fields such as the dark gauge boson that is studied in the present paper.  An important question is whether we can detect these axion-like fields when they do not interact directly with the visible sector. Since any type of matter is universally coupled to gravity according to the equivalence principle, gravitational waves are the natural (if not the only) candidates to probe these axion-like fields. More precisely, in our setup, the amplified dark photons serve as a quadratic source for the linear equations of motion of the tensor perturbations leading to the production of gravitational waves. Due to the parity-violating nature of the Chern-Simons interaction, one helicity of tensor modes is more amplified than the other, leading to the chiral gravitational waves. This scenario was recently suggested in Refs. \cite{Machado:2018nqk,Machado:2019xuc} and analyzed by means of numerical methods. In this paper, we have studied this scenario analytically. 

Section \ref{sec-vector-field} is devoted to the analytic study of dark photon production from axion-like field. Ignoring the nonlinear effects, such as backreaction and backscattering, we have studied the amplification of the dark photons due to the tachyonic and also semi-tachyonic resonances. Using the methods that had been adopted to study the non-perturbative preheating process after inflation, we have found an analytic expression for the number density of the produced particles in terms of the mass and coupling of the axion. We have also studied the effects of backreaction and backscattering and estimated the time when these effects become important and beyond which the linear analysis cannot be trusted anymore. Further, for the range of parameters where these effects become significant, we have obtained an analytic expression for the number density of dark photons. With this assumption we have obtained an analytic estimation for the correction to the effective number of the relativistic degrees of freedom induced by the produced dark photons. Comparison with numerical results of Ref.~\cite{Machado:2018nqk} shows the consistency of our result within an $\order{1}$ prefactor. 

In section \ref{sec-GWs} we looked for the effects of amplified dark photons on the gravitational wave spectral density. Since the tachyonic resonance of the dark photon is the source of the tensor modes, it is not easy to solve the equations of motion for the gravitational waves analytically and that is the reason why this problem was studied numerically in the previous works. Using the saddle point approximations, we have found that there are many saddle points that may play roles in the integral of the dark photon contributions to the gravitational waves. We have shown that most of these saddle points neatly cancel one another so that we only need to take into account the effects of the most important saddle point. In this regard, we have found an analytical expression for the peak of the gravitational wave spectral energy density in terms of the parameters of the model. We have compared the analytic result for the gravitational wave spectrum with numerical simulations of Ref.~\cite{Machado:2018nqk} and showed that they are in a very good agreement.

The analytic results of our paper, besides giving insight to the process of particle production of the model, are useful for observational purposes. The analytic expression for the gravitational wave spectrum can be used as a template to extract signal from observations. By fitting to the data one can find the amplitude and the position of the peak which are directly related to the parameters of the model. Further, one can in principle distinguish between other gravitational wave production mechanisms by checking consistency of other features such as the chirality of gravitational waves or the change in the effective relativistic degrees of freedom.

\subsection*{Supplementary note}
After completion of this work, the paper \cite{Ratzinger:2020oct} appeared where the authors have performed precise lattice simulations of the model, i.e. they have solved for the coupled system of axion, dark photon and metric fluctuations. Their numerical analyses confirm our analytical results at the linear regime. On the other hand, new features are found at the nonlinear regime when the backreaction and backscattering effects become important. First, the axion remnant decays with slower rate in comparison with the linear analysis result. Second, the chirality of the gravitational waves can be washed out for some allowed regions of the parameter space.

\subsection*{Acknowledgments}
We thank A.A. Abolhasani, T. Fujita, F. Hajkarim, S. Hooshangi, M.H. Namjoo, M. Noorbala, and S. Shakeri for insightful discussions. B.S. thanks Yukawa Institute for Theoretical Physics at Kyoto university for their hospitality during the time this work was initiated and was in progress. B.S. thanks A. Mohammadi for help on the figures. The work of M.A.G. was supported by Japan Society for the Promotion of Science Grants-in-Aid for international research fellow No. 19F19313. The work of S.M. was supported in part by Japan Society for the Promotion of Science Grants-in-Aid for Scientific Research No.~17H02890, No.~17H06359, and by World Premier International Research Center Initiative, MEXT, Japan. 
 
\appendix

\section{Polarization vectors and tensors}\label{app:pol}

In this appendix we summarize some formulas about the polarization vectors and tensors which are used throughout the paper. For the vector field we have chosen the Coulomb gauge thus $A_i$ is orthogonal to $k_i$ and can be expressed as a combination of linear polarization vectors $\varepsilon_i^{(a)}(\kv)$ for $a=1,2$ which satisfy 
\eq{
\varepsilon_i^{(a)}\varepsilon_i^{(b)}=\delta_{ab}\,,\qquad\epsilon_{ij\ell}\varepsilon_j^{(1)}\varepsilon_\ell^{(2)}=\frac{k_i}{k}\,,
}
where we have omitted the dependence of polarization vectors on $\kv$ for brevity. Circular polarization vectors can be constructed as 
\eq{
\varepsilon_i^\lambda=\frac{1}{\sqrt{2}} 
\Big( \varepsilon_i^{(1)}+i\lambda \varepsilon_i^{(2)} \Big) \,,
}
where $\lambda$ can be $+1$ or $-1$ corresponding to right- or left-handed circular polarization. Then it is easy to see that 
\eqa{
\label{eq:id1}
k_i\varepsilon_i^\lambda=0\,,&\qquad
\varepsilon_i^\lambda\varepsilon_i^{\lambda'}{}^*=\delta_{\lambda\lambda'}\,,\\
\label{eq:id2}
\epsilon_{ij\ell}\varepsilon_j^\lambda\varepsilon_\ell^{\lambda'}{}^*=-i\lambda\delta_{\lambda\lambda'}\frac{k_i}{k}\,,&\qquad
\epsilon_{ij\ell}k_j\varepsilon_\ell^\lambda=-i\lambda\varepsilon_i^\lambda k \,,\\
\label{eq:id3}
\sum_{\lambda=\pm}\varepsilon_i^\lambda\varepsilon_j^\lambda&=\delta_{ij}-\frac{k_ik_j}{k^2}\,.
}  
Note that $\varepsilon_i^{\lambda}(\kv){}^*=\varepsilon_i^{-\lambda}(\kv)=\varepsilon_i^{\lambda}(-\kv)$. A useful identity for polarization vectors corresponding to different directions is
\eq{
\label{epsilon-identity}
\varepsilon_i^{\lambda}(\kv)\varepsilon_i^{\lambda'}(\kpv){}^*
= \frac{1}{2} ( 1 + \lambda\lambda' \cos\gamma ) \,,
}
where $\gamma=\cos^{-1}({k_ik'_i}/{kk'})$ is the angle between the two vectors $\kv$ and $\kpv$. Similarly we can construct linear polarization tensors
\eqa{
e_{ij}^{p}&=\varepsilon_i^{(1)}\varepsilon_j^{(1)}-\varepsilon_i^{(2)}\varepsilon_j^{(2)} \,, \\
e_{ij}^{c}&=\varepsilon_i^{(1)}\varepsilon_j^{(2)}+\varepsilon_i^{(2)}\varepsilon_j^{(1)} \,,
}
where superscripts $p$ and $c$ correspond to plus and cross polarizations. Furthermore, circular polarization tensor is defined to be
\eq{
e_{ij}^\lambda=\frac{1}{2}(e_{ij}^{p}+i\lambda e_{ij}^{c})\,,
}
with $\lambda=\pm1$. A useful relation between polarization vector and tensor is the following
\eq{
e_{ij}^\lambda=\varepsilon_i^{\lambda}\varepsilon_j^{\lambda}\,,
}  
such that all identities of Eqs.~\eqref{eq:id1}-\eqref{eq:id3} can be translated straightforwardly  to the case of tensors.  
\section{Basic Floquet theory}\label{app:floquet}
In this appendix we review the classification of the solutions of a linear differential equation with periodic coefficients, namely the Floquet theory. For more details see for example Refs.~\cite{McLachlan:1964,Grimshaw:1991}. If we have the  linear system of differential equations 
\eq{
\label{eq:mateq}
\dot{\xr}={\rm A}(t) \, \xr\,,
}
with ${\rm A}(t)$ a $T$-periodic $n\times n$ matrix and $\xr$ a $n\times 1$ vector, then a typical solution $\xr(t)$ does not have to be periodic but instead is of the form
\eq{
\xr(t)=e^{\mu t}p(t)\,,
}
where $p(t)$ is a periodic function with period $T$ and $\mu$ is one of the $n$ characteristic exponents $\mu_i$ which are \emph{intrinsic} to the problem and satisfy
\eq{
\label{eq:trA}
e^{\mu_i T}\dots e^{\mu_n T}=\exp(\int_{0}^{T}\dd{t'}\tr({\rm A}(t')))\,.
}
More specifically, if we have $n$ linearly independent solutions $\xr^{(n)}(t)$ of the problem, we construct the principal fundamental matrix $\vb{X}(t)$ 
\eq{
\label{eq:Xt}
\vb{X}(t)=\left[\xr^{(1)}(t),\dots,\xr^{(n)}(t)\right]\,,
}
such that $\vb{X}(0)=I$. Then $\rho_i\equiv\exp(\mu_i T)$ are eigenvalues of $B\equiv\vb{X}(T)$ and $\rho_i$ are sometimes called characteristic multipliers. Note that we have three categories
\begin{itemize}
\item $|\rho|<1$ which means $\Re(\mu)<0$ then $\xr(t)$ tends to zero at infinity.
\item $|\rho|=1$ which means $\Re(\mu)=0$ then the solution is bounded. If $\rho=\pm1$ the solution is periodic.
\item $|\rho|>1$ which means $\Re(\mu)>0$ then $\xr(t)$ goes to infinity as $t\to\infty$.    
\end{itemize}
The special interesting case for us is a second order linear differential equation of the form 
\eq{
\label{eq:diffeq}
\ddot{x}+a(t)x=0\,,
}
where $a(t)$ is $T$-periodic. It can be put into the general form of Eq.~\eqref{eq:mateq} if we define
\eq{
\xr=
\begin{bmatrix}
x\\\dot{x}
\end{bmatrix}\qquad
{\rm A} =
\begin{bmatrix}
0&\phantom{-}1\\
-a(t)&\phantom{-}0
\end{bmatrix}\,.
}
Then we can find two linearly independent solutions with linearly independent initial conditions
\eq{
\xr^{(1)}(0)=
\begin{bmatrix}
1\\0
\end{bmatrix}\,,\qquad
\xr^{(2)}(0)=
\begin{bmatrix}
0\\1
\end{bmatrix}\,.
}
Form Eq.~\eqref{eq:trA} and definition of $\rho_i$ we have
\eqa{
\rho_1\rho_2&=1 \, ,\\
\rho_1+\rho_2&=\tr(B)=x^{(1)}(T)+\dot{x}^{(2)}(T)\,.
}
If we define $\theta\equiv\tr(B)/2$ then we have
\eq{
\rho_{1,2}=\theta\pm\sqrt{\theta^2-1}\,.
}
In terms of characteristic exponents $\mu_1$ and $\mu_2$ we find out that $\mu_1+\mu_2=0$ and $\cosh(\mu_1T)=\theta$. Then it is easy to see that the solution is stable for $-1<\theta<1$ and unstable if $|\theta|>1$. The boundary between stable and unstable solution is given by the condition $\theta=1$ or $\theta=-1$ which correspond to a $T$-periodic or $2T$-periodic solutions respectively \cite{Grimshaw:1991}. Thus we can find boundaries, in the parameter space of the problem, between stable and unstable solutions by making an ansatz which is $T$-periodic 
\eq{
\label{eq:T}
x(t)=\sum_{n=0}^{\infty}a_n\cos(\frac{2\pi n}{T}t)+\sum_{n=1}^{\infty}b_n\sin(\frac{2\pi n}{T}t)\,,
}
or $2T$-periodic
\eq{
\label{eq:2T}
x(t)=\sum_{n=0}^{\infty}a_n\cos(\frac{\pi n}{T}t)+\sum_{n=1}^{\infty}b_n\sin(\frac{\pi n}{T}t)\,,
}
and demand that they be a solution of Eq.~\eqref{eq:diffeq}. This gives us four homogeneous sets of linear system of equations with $a_n$s and $b_n$s in Eqs.~\eqref{eq:T} and \eqref{eq:2T} as unknowns. A nonzero solution for $a_n$s or $b_n$s is achieved if the determinant of the coefficient matrix is zero. This gives us four different relations among parameters of the differential equation. The coefficient matrix is indeed infinite dimensional but we can approximately solve these equations by keeping only a $N\times N$ sub-matrix for large enough $N$.

The interesting second order equation for us is the Mathieu equation $\ddot{x}+(A_{\rm m}-2q_{\rm m}\cos(2t))x=0$ in which $T=\pi$. In the limit that $q_{\rm m}\ll{A}_{\rm m}$ such that we can ignore the oscillating term, we can approximately construct the matrix $B$ defined below Eq.~\eqref{eq:Xt} as follows 
\eq{
B=
\begin{bmatrix}
\cos(\pi\sqrt{A_{\rm m}})&\frac{1}{\sqrt{A_{\rm m}}}\sin(\pi\sqrt{A_{\rm m}})\\
-\sqrt{A_{\rm m}}\sin(\pi\sqrt{A_{\rm m}})&\cos(\pi\sqrt{A_{\rm m}})
\end{bmatrix}\,.
} 
As a result, we have $\theta=\cos(\pi\sqrt{A_{\rm m}})$. Then we must have $A_{\rm m}=(2n)^2$ for $\theta=1$ and $A_{\rm m}=(2n-1)^2$ for $\theta=-1$ with $n=1,2,\dots$. We can find periodic solutions as perturbative series in $q_{\rm m}$
\eqa{
x(t)&=x_0+q_{\rm m} x_1+\dots\\
A_{\rm m}&=n^2+q_{\rm m}A_{\rm m}^{(1)}+\dots\,,
}
for $n=1,2,\dots$. Following this analysis it can be shown that for the first instability band $n=1$, we must have 
\eq{
\label{eq:band1}
1-q_{\rm m}<A_{\rm m}<1+q_{\rm m}\,,
}
while for higher $n$ the width of the instability band is of order $\order{q_{\rm m}^n}$ \cite{landau:mech}. Also in the first band the characteristic exponent is approximately given by
\eq{
\mu=\frac{1}{2}\sqrt{q_{\rm m}^2-(A_{\rm m}-1)^2}\,.
}
For larger $q_{\rm m}$ there is no closed form solution and we must solve the system numerically to find the characteristic exponents. The boundaries between the stable and unstable solutions can be found by the ansatz in Eqs.~\eqref{eq:T} and \eqref{eq:2T}. For Eq.~\eqref{eq:matheiu} this procedure results in the stability/instability chart shown in Fig.~\ref{fig:chart}.

\section{Semi-tachyonic regime}\label{semitach}
In this appendix we derive the transfer matrix in the semi-tachyonic regime. Around $x_j^{\rm min}$ (see shaded regions after $x_{\rm et}$ in Fig. \ref{fig:delta}), the equation for the mode function Eq.~\eqref{eq:chiy} can be approximately written as
\eq{
	\dv[2]{\chi}{\tau}+(\tau^2-\vrk^2)\chi=0\,,
} 
where we have defined $\tau\equiv (q_j/2)^{1/4}(x-x_j^{\rm min})$ and $\vrk^2$ is given by Eq.~\eqref{eq:adappmin}. The general solution is given by
\eq{
	\label{eq:Dnu}
	\chi=c_1D_{\nu_+}((1+i)\tau)+c_2D_{\nu_-}((-1+i)\tau)\,,
}
where $\nu_\pm \equiv -(1\pm i\vrk^2)/2$ and $D_\nu(\tau)$ is the parabolic cylinder function which is the solution of the standard differential equation $y''+(\nu+1/2-\tau^2/4)y=0$ (see Ref.~\cite{NIST:DLMF} for more details) with $c_1$ and $c_2$ are some constants. Far from the point $\tau=0$ (or equivalently $x=x_j^{\rm max}$), in the region where $\omega^2>0$, adiabatic approximation is valid and Eq.~\eqref{eq:posmass} can be used. However, there is an overlapping region which is not too close to the point $\tau=0$ to break the adiabatic approximation and still not too far so that the solution of the form \eqref{eq:Dnu} is still valid. The asymptotic form of the parabolic cylinder functions for $\tau>0$ can be obtained to be
\eq{
	\spl{
		D_\nu((1+i)\tau)&\to2^{\nu/2}e^{i\pi\nu/4}\frac{e^{-i\tau^2/2}}{\tau^{-\nu}} \,, \\ D_\nu((-1+i)\tau)&\to2^{\nu/2}e^{3i\pi\nu/4}\frac{e^{i\tau^2/2}}{\tau^{-\nu}}+\frac{\sqrt{2\pi}}{(1-i)^{1+\nu}\Gamma(-\nu)}\frac{e^{-i\tau^2/2}}{\tau^{1+\nu}}\,,
	}
}
and for $\tau<0$
\eq{
	\spl{
		D_\nu((1+i)\tau)&\to2^{\nu/2}e^{i\pi\nu/4}\frac{e^{-i\tau^2/2}}{\tau^{-\nu}}-\frac{(1-i)e^{-5i\pi\nu/4}\sqrt{\pi}}{2^{(1+\nu)/2}\Gamma(-\nu)}\frac{e^{i\tau^2/2}}{\tau^{1+\nu}} \,, \\
		D_\nu((-1+i)\tau)&\to(1+i)^{\nu}e^{i\pi\nu/2}\frac{e^{i\tau^2/2}}{\tau^{-\nu}}\,.
	}
}
In the overlapping region, we can write the exponent in Eq.~\eqref{eq:posmass} as
\eq{
	\int\dd{x}\omega\approx\int\dd{\tau}\sqrt{\tau^2-\vrk^2}=\frac{1}{2}\left[\tau\sqrt{\tau^2-\vrk^2}-\vrk^2\log(\tau+\sqrt{\tau^2-\vrk^2})\right]\,.
}
Then in the limit $|\tau|\gg\vrk$ and for $\tau<0$ we can write Eq.~\eqref{eq:posmass} as 
\eq{
	\label{eq:chij}
	\chi_j\approx\frac{1}{(q_j/2)^{1/8}\sqrt{-2\tau}}\left[\alpha_j e^{-i\theta'_j}\frac{e^{i\tau^2/2}}{(-\tau)^{i\vrk^2/2}}+\beta_j e^{i\theta'_j}\frac{e^{-i\tau^2/2}}{(-\tau)^{-i\vrk^2/2}}\right]\,,
}
where $\theta'_j\equiv \theta_j+\frac{\vrk^2}{4}\left(1+\log(\frac{4}{\vrk^2})\right)$ and $\theta_j$ is the accumulated phase by the time $\tau=0$ defined below Eq.~\eqref{eq:Xjdef}. Similarly for $\tau>0$ we write
\eq{
	\label{eq:chij1}
	\chi_{j+1}\approx\frac{1}{(q_j/2)^{1/8}\sqrt{2\tau}}\left[\alpha_{j+1} e^{-i\theta''_j}\frac{e^{-i\tau^2/2}}{\tau^{-i\vrk^2/2}}+\beta_{j+1} e^{i\theta''_j}\frac{e^{i\tau^2/2}}{\tau^{i\vrk^2/2}}\right]\,,
}
where $\theta''_j \equiv \theta_j-\frac{\vrk^2}{4}\left(1+\log(\frac{4}{\vrk^2})\right)$. The solutions of Eqs.~\eqref{eq:chij} and \eqref{eq:chij1} are connected via Eq.~\eqref{eq:Dnu} and we can obtain a transfer matrix to relate the coefficients $(\alpha_{j},\beta_{j})$ to $(\alpha_{j+1},\beta_{j+1})$ as
\eq{
	\label{eq:tmat}
	\begin{bmatrix}
		\alpha_{j+1}\\\beta_{j+1}
	\end{bmatrix}
	=
	\begin{bmatrix}
		e^{-i\vartheta}\sqrt{1+e^{\pi\vrk^2}}&ie^{\pi\vrk^2/2}e^{2i\theta_j}\\
		-ie^{\pi\vrk^2/2}e^{-2i\theta_j}&e^{i\vartheta}\sqrt{1+e^{\pi\vrk^2}}
	\end{bmatrix}
	\begin{bmatrix}
		\alpha_{j}\\\beta_{j}
	\end{bmatrix}\,,
}
where 
\eq{
	\label{eq:vartheta}
	\vartheta \equiv \arg\Gamma\left(\frac{1}{2}+i\frac{\vrk^2}{2}\right)+\frac{\vrk^2}{2}\left(1+\log(\frac{2}{\vrk^2})\right)\,,
}
in which we have used the fact that $\Big|\Gamma\left(\frac{1}{2}+i\frac{\vrk^2}{2}\right)\Big|^2=\frac{\pi}{\cosh(\pi\vrk^2/2)}$. The transfer matrix in Eq.~\eqref{eq:tmat} is the counterpart of Eq.~\eqref{eq:mattach} obtained in the tachyonic regime. Note that the transfer matrix is the same as obtained in  Ref.~\cite{Kofman:1997yn} if we replace $\kappa^2$ in their expression by $-\vrk^2$ everywhere except in the Logarithm of Eq.~\eqref{eq:vartheta}. The transfer matrix Eq.~\eqref{eq:tmat} is used to obtain a recursive relation for number of dark photons in the main text. 

\section{Possible perturbative decay}\label{app:perturbative-decay}

The non-perturbative particle production that we studied in section \ref{sec-vector-field} is all one needs to study production of the GWs by dark photons in our setup. However, as we mentioned at the end of subsection \ref{subsec-mode-function}, the subsequent evolution of the remnant of the axion with energy density $\rho^{\rm rem}_{\bar\phi}$ after the end of the non-perturbative particle production is important to see whether axion remnant decays or survives. Indeed, depending on the values of the mass and coupling, the decay is possible. If axion remnant decays, it perturbatively produces again dark photons which contribute to the effective number of relativistic degrees of freedom. Otherwise, it survives and would contribute to the dark matter as it is well known in the context of the axionic vector dark matter models. In this appendix, we track analytically the evolution of the axion remnant after the non-perturbative particle production.

\subsection{Decay rate of axion to photons}

Let us first compute the axion decay rate $\Gamma_{{\bar\phi}\to{AA}}$ to the photons. Of course we know the answer from the standard perturbative analysis of quantum field theory. However, here we find it via the adiabatic approximation method that we presented in section \ref{sec-vector-field} and through the Boltzmann equation following the method suggested in Ref. \cite{Braden:2010wd}. The readers who are not interested in this derivation can simply move to the next subsection.

Since we are in the perturbative regime, we can iteratively solve Eq. (\ref{eq:alpha-beta}) for $|{\bar \beta}|\ll1$ with the initial conditions ${\bar \alpha}(x=x_{\rm ep})=1$ and ${\bar \beta}(x=x_{\rm ep})=0$ which yields
\begin{equation}\label{beta-pert-def}
\bar{\beta}(x) \approx \frac{1}{2} \int_{x_{\rm ep}}^x \frac{d{\omega}(x')}{\omega(x')} 
e^{-2i\int dx'' \omega(x'')} \,,
\end{equation}
where the notation ${\bar \alpha}$ and ${\bar \beta}$ show that they correspond to the perturbative particle production. We use this notation for the number density and energy density of the particles that produce through the perturbative decay as well. Moreover, the initial condition ${\bar \beta}(x=x_{\rm br})=0$ corresponds to the state without any particle while we know that some dark photons are already produced through the non-perturbative particle production process. By considering ${\bar \beta}(x=x_{\rm br})=0$ and the notation ${\bar \beta}$ used in Eq. (\ref{beta-pert-def}) we mean that we only look at particles that are produced through the perturbative process. 
In this regime, we approximately have $e^{-2i\int dx'' \omega(t'')} \approx e^{-2i\kappa\int dx''/a(x'')}$ and also from (\ref{eq:chiy}) we find
\begin{equation}
\frac{1}{\omega} \frac{d{\omega}}{dx} \approx 
- \frac{\lambda\varphi}{2\kappa} \frac{\cos{x}}{\sqrt{a}} \,,
\end{equation}
where we have also neglected time derivative of the scale factor as before. 

Substituting the above results in (\ref{beta-pert-def}) yields
\begin{equation}\label{beta-pert}
{\bar \beta}(x) \approx -\frac{\lambda\varphi}{8\kappa} 
\int_{x_{\rm ep}}^x \frac{dx'}{\sqrt{a(x')}} 
\Big( e^{i\kappa\psi_+} - e^{-i\kappa\psi_-}\Big) \,; \hspace{1cm} 
\psi_{\pm} \equiv \mp 2 \int \frac{dx''}{a(x'')} + \frac{x'}{\kappa} \,.
\end{equation}

The above oscillatory integral can be approximated by the method of saddle point approximation. The saddle happens at the time
\begin{equation}\label{saddle-point}
\frac{d\psi_+}{dx}(x=x_{\rm s}) = 0 \, \Rightarrow a_{\rm s} = 2 \kappa \,,
\end{equation}
where $a_{\rm s} = a(x=x_{\rm s})$ is the scale factor at the saddle time $x_{\rm s}$. Using this result, we find the number density of dark photons during perturbative decay of axion as follows
\begin{equation}\label{n-pert}
\bar{n}_{k,\lambda} = |{\bar \beta}|^2 
= \frac{\pi {m} \varphi^2}{64 \kappa^3 H_{\rm s}} \,,
\end{equation}
where $H_{\rm s} = H(x=x_{\rm s})$ is the value of the Hubble expansion rate at the saddle time. The above expression gives the number of produced particles through the perturbative decay of the axion to the dark photons. 

Ignoring the vacuum energy density, the energy density of the dark photons ${\bar \rho_{A}}$ that are produced  perturbatively is given by
\begin{equation}\label{rho-darkphoton-B-def}
a^4 {\bar \rho_{A}} \approx \sum_{\lambda=\pm} \int \frac{d^3k}{(2\pi)^3} 
k\, {\bar n}_{k,\lambda} \,,
\end{equation}
where we have used $\omega_k \approx k/a$ to the first order of approximation in this regime. Substituting from (\ref{n-pert}), we find
\begin{equation}\label{rho-darkphoton-B}
a^4 {\bar \rho}_{A} \approx \frac{\varphi^2 m^5}{64\pi}
\int_{\frac{a_{\rm ep}}{2}}^{\frac{a}{2}} \frac{d\kappa}{H_{\rm s}} \,.
\end{equation}

Having this, we can find the decay rate after the time $x_{\rm ep}$ from the definition of the Boltzmann equation
\begin{equation}\label{Boltzmann-Eq}
\frac{1}{a^4}\frac{d}{dt} \Big(a^4 {\bar \rho}_{A}\Big) = \Gamma_{{\bar\phi}\to{AA}} \, \rho^{\rm rem}_{\bar\phi} \,,
\end{equation}
where $\Gamma_{{\bar\phi}\to{AA}}$ is the decay rate and $\rho^{\rm rem}_{\bar\phi}$ is the energy density of the remnant of the axion. Taking the time derivative of (\ref{rho-darkphoton-B}) and then comparing the result with the Boltzmann equation (\ref{Boltzmann-Eq}), we find the following expression for the decay rate of the axion remnant
\begin{equation}\label{Gamma}
\Gamma_{{\bar\phi}\to{AA}} = \frac{{\alpha}_{a}^2 m^3}{64\pi \f^2} \,,
\end{equation}
which is in agreement with the standard result of quantum field theory for a decay through a trilinear interaction. We use this result to find the time scale of the decay of the axion remnant in the next subsection.

\subsection{Fate of the axion remnant}

As we already mentioned, some parts of the energy density of the initial homogeneous axion field are transferred to the dark photons during the non-perturbative process of tachyonic and also possibly semi-tachyonic and parametric resonances while the remaining part is the remnant of the axion which is dealt with in this appendix. 

During the nonperturbative production of the dark photons, the energy conservation equation between the axion field and dark photons can be written in the form
\begin{equation}\label{CoE}
\frac{1}{a^3}\frac{d(\rho_{\bar\phi} a^3)}{dt} 
= - \frac{1}{a^{4}} \frac{d(\rho_A a^4)}{dt} \,,
\end{equation}
where $\rho_{\bar\phi}$ is the energy density of the initial homogeneous axion field and $\rho_A$ is the energy density of dark photon produced  non-perturbatively  through the resonance processes. The interaction between the axion field and the dark photons is given by the Chern-Simons term $\phi{F}{\tilde F}$. Non-perturbative particle production starts from $a_{\rm os}$ and ends at $\min \left[a_{\mathrm{br}}, a_{\mathrm{en}}\right]$. Therefore, we should integrate the above equation from $x_{\rm os}$ to $\min \left[x_{\mathrm{br}}, x_{\mathrm{en}}\right]$. As we already mentioned, for the favourable range of parameter space, backreaction terminates the process of non-perturbative particle production and therefore we assume that non-perturbative particle production happens in a very short time interval and $\min \left[a_{\mathrm{br}}, a_{\mathrm{en}}\right]=a_{\mathrm{br}}$. In that case, the above equation can be integrated approximately to find the axion remnant at the time of end of non-perturbative particle production as 
\begin{equation}
\rho^{\rm rem}_{{\bar\phi}}(a_{\rm br}) \lesssim \frac{m^2 \phi_{\rm os}^2}{2 a_{\rm br}^3} - \rho_{A}(a_{\rm br}) \,,
\end{equation}
where the inequality appears when we take into account the effects of backreaction and backscattering. If the time duration $[x_{\rm os},x_{\rm br}]$ is large so that $\frac{a_{\rm br}}{a_{\rm os}} \sim {\cal O}(10)$, the above approximation is no longer applicable and we need to solve Eq. (\ref{CoE}) numerically\footnote{For instance, in the case of QCD axions, numerical analysis shows that backreaction becomes important very soon for $\frac{a_{\rm br}}{a_{\rm os}} \sim {\cal O}(10)$ \cite{Agrawal:2017eqm}.}. However, as we will see, we do not need the explicit form of the remnant and we determined it just for the concreteness. After the tachyonic resonance terminates at the time $x_{\rm br}$, we are left with the axion remnant which approximately given by $\rho^{\rm rem}_{{\bar\phi}}(a_{\rm br})$ and also dark photons $\rho_A$ that are produced via non-perturbative resonance process.

The decay rate of the axion to the dark photons is given by \eqref{Gamma}. After non-perturbative particle production is terminated by the backreaction, the Hubble expansion rate decreases as $H^2\sim a^{-4}$ in the RD era and, depending on the mass and coupling, it can approach the axion decay rate (\ref{Gamma}). When the Hubble expansion rate drops below the decay rate \eqref{Gamma}, we would have perturbative particle production for the remnant of the axion. This is similar to what happens for the inflaton during perturbative reheating after non-perturbative preheating. From the conservation of the energy after the time $x_{\rm br}$, we have 
\begin{equation}\label{Boltzmann-Eq-phi}
\frac{d}{dt} \Big(a^3 \rho^{\rm rem}_{{\bar\phi}}\Big) = 
- \Gamma_{{\bar\phi}\to{AA}} \big(a^3  \rho^{\rm rem}_{{\bar\phi}} \big) \, \hspace{.5cm}
\Rightarrow \hspace{.5cm} a^3 \rho^{\rm rem}_{{\bar\phi}} 
\sim e^{-\Gamma_{{\bar\phi}\to{AA}}t} \,.
\end{equation}

The above result shows that the axion remnant will decay completely to the dark photons in time scale $t_{\rm decay} \sim \Gamma_{{\bar\phi}\to{AA}}^{-1}$. To have axionic dark matter, we need to prevent the axion to decay, requiring $t_{\rm decay}>H^{-1}$ or equivalently
\begin{equation}\label{decay-condition}
\Gamma_{{\bar\phi}\to{AA}} < H \,.
\end{equation}
Therefore, from Eq. (\ref{Gamma}) we see that depending on the mass and coupling, the above condition may or may not be satisfied.

In summary, the axion mass determines the time that axion starts to oscillate while the combination of the mass and coupling determines the decay time. If the mass and coupling are such that the axion remnant survives, it can contribute to the dark matter  \cite{Agrawal:2017eqm,Agrawal:2018vin,Dror:2018pdh,Co:2018lka,Bastero-Gil:2018uel}. This is the necessary condition for the axion dark matter scenarios but it is not enough. For example, depending on the parameters, the remnant can be considered as a part of dark matter or possibly the whole dark matter. In the latter case, there is a low mass bound $m \geq 10^{-22}$ eV \cite{Hu:2000ke,Arvanitaki:2009fg,Fan:2016rda}. Otherwise, it can decay into the dark photons which contribute to the number of relativistic degrees of freedom similar to what we considered for the dark photons produced non-perturbatively  in subsection \ref{N-eff}. In this case, axion decays completely in the short time scale $\Gamma^{-1}_{{\bar\phi}\to{AA}}$, as shown in (\ref{Boltzmann-Eq-phi}), and we can approximately consider instantaneous decay. Therefore, the whole of axion remnant energy density is converted to the dark photons through instantaneous perturbative decay around the time $x_{\rm dec}$ when $H(x_{\rm dec}) \sim \Gamma_{{\bar\phi}\to{AA}}$. The shift in the effective number of relativistic degrees of freedom then will be
\begin{equation}\label{N-eff-NP-P}
\Delta{N}_{\rm eff}|_{t_{\rm dec}} 
\approx \frac{8}{7}\Big(\frac{11}{4}\Big)^{\frac{4}{3}} \frac{\rho^{\rm rem}_{{\bar\phi}}(a_{\rm dec})}{\rho_\gamma(a_{\rm dec})} \,,
\end{equation}
and the weakest condition that has to be satisfied  in order  to respect the predictions of the standard big bang cosmology is $\Delta {N}_{\rm eff}|_{t_{\rm dec}} < {\cal O}(1)$.


\bibliography{references} 
\bibliographystyle{JHEP}
\end{document}